\documentclass[preprint]{aastex}
\def\simgr{\,\hbox{\hbox{$ > $}\kern -0.8em \lower 1.0ex\hbox{$\sim$}}\,}
\def\simle{\,\hbox{\hbox{$ < $}\kern -0.8em \lower 1.0ex\hbox{$\sim$}}\,}

\shortauthors{THORSTENSEN et al.}
\shorttitle{New Cataclysmics}

\begin{document}
\title{A Trip to the Cataclysmic Binary Zoo: Detailed Follow-Up of 35 
Recently-Discovered Systems
}

\author{John R. Thorstensen, Erek H. Alper, and Kathryn E. Weil
}

\affil{Department of Physics and Astronomy\\
6127 Wilder Laboratory, Dartmouth College\\
Hanover, NH 03755-3528}

\begin{abstract}
We report follow-up studies of 35 recently-discovered cataclysmic
variables (CVs),  32 of which were found in large, automated
synoptic sky surveys.  The objects were selected for observational
tractability.  For 34 of the objects we 
present mean spectra and spectroscopic orbital periods, and for
one more we give an eclipse-based period. Thirty-two
of the period determinations are new, and three of these
refine published estimates based on superhump periods.
The remaining three of our determinations confirm previously 
published periods.
Twenty of the stars are confirmed or suspected dwarf novae with 
periods shorter than 3 hours, but we also find three apparent 
polars (AM Her stars), and six systems with $P > 5$ h, 
five of which have secondary stars visible in their spectra, from
which we estimate distances when possible.
The orbital period distribution of this sample is 
very similar to that of previously discovered CVs.
\end{abstract}


\section{Introduction}

Cataclysmic variable stars (CVs) are a broad class of binaries
that include a white dwarf primary star accreting via 
Roche lobe overflow from a close companion, which usually
resembles a cool main-sequence star; 
\citet{warner95} gives a useful introduction.

CVs are compact, low-luminosity systems with 
typical orbital periods $P_{\rm orb}$ of only a few 
hours.  Their evolution is driven largely by 
angular momentum loss, which over most of a
CV's lifetime causes the orbit to gradually
shrink.  However, at a period
near 75 min, the slope of the secondary's  
mass-radius relation reverses (at least for
hydrogen-burning secondaries) which causes the orbital
period to increase as angular momentum is lost. 

Single white dwarfs evolve from red giants, 
which attain radii 
much larger than typical CV systems.  Unless
a white dwarf is formed elsewhere and later captured --
an unlikely event in the field -- a CV must
therefore have passed through common-envelope
evolution at some point, during which the 
secondary was engulfed by the primary, leading
to rapid loss of angular momentum.  Modeling
the common-envelope stage with {\it a priori} physics is 
extremely difficult, so the usual practice is to
use simple parameterizations to describe the process,
and match model outputs to the observed population
(see, e.g., \citealt{goliaschnelson15}).

Constraining the calculations requires an accurate
accounting of the CV population.  
CVs are rather rare, low-luminosity systems, so 
this requires consideration of the channels through
which they are discovered.

The first CVs were discovered because they were
variable stars, most conspicuously classical novae, 
driven by thermonuclear explosions of hydrogen-rich
material accreted onto the white dwarf, and also
dwarf novae, in which the accretion
disk around the white dwarf transitions from time to 
time into a much brighter state.  Other CVs with
less dramatic variability
were found by color selection, especially ultraviolet 
excesses.  The Palomar-Green survey 
found a sizable number of CVs
\citep{green82, ringwald93},
and more recently the Sloan Digital Sky Survey (SDSS)
obtained spectra of 285 color-selected CVs and
CV candidates, 
most of which were new discoveries 
(see \citealt{szkodyviii}
and previous papers in that series).
Finally, many CVs emit X-rays; X-ray surveys especially
tend to find CVs in which the white-dwarf primary
is strongly magnetized (see, e.g., \citealt{thorhalpern13,
halpernthor15}).  

In recent years several synoptic sky surveys have 
generated a torrent of new CV discoveries. 
The Catalina Real Time Transient Surveys (CRTS; \citealt{drakecrtts, 
breedt14, drake14}) have discovered over 1000 CVs on
the basis of variability, most of them being discovered
by the Catalina telescope and denoted as ``CSS''.  The Russian MASTER project
\citep{lipunov10} has also produced a large number of 
new discoveries, and most recently the ASAS-SN 
project \citep{shappeecurtain}, has come into its own.
ASAS-SN uses 14-cm aperture lenses and reaches
a limiting magnitude of only $\sim 16$, but is now covering
$\sim 20,000$ square degrees per night under favorable
conditions (K. Stanek, private communication). 
Many transient sources fade back to magnitudes
beyond the reach of spectroscopic followup on 
2-m class telescopes, but a substantial 
minority remain tractable.  ASAS-SN's fast cadence
and relatively shallow limiting magnitude make it an 
especially rich source of CVs amenable to further
study.

CVs also continue to be discovered through color 
selection.  As examples, \citet{carter13} searched the SDSS
for AM CVn stars (ultra-short period CVs in 
which both components are degenerate dwarfs),
and \citet{kepler16} searched for hot white dwarfs
and turned up twelve CVs as a byproduct.

To place the many newly-discovered CVs in context, it 
is necessary to follow them up, in particular to find
orbital periods.  Periods can often be found from
photometry, but in non-eclipsing systems radial-velocity 
spectroscopy 
gives more definitive results. 
Spectroscopy also provides a wealth of ancillary information.
We present here follow-up observations, mostly spectroscopic,
of 35 newly-discovered CVs, 32 of which were found by 
CRTS, ASAS-SN, and/or MASTER.  For all,
we determine orbital periods.  Thirty-two
of the 35 periods are apparently are not published elsewhere,
though superhump periods
were available for three of of the objects.  For three
other objects we confirm periods that have appeared in the 
literature. 

The objects studied here were selected mostly on the
basis of their observational tractability, and the 
lack of a published orbital period (although in three cases
we became aware of published periods before submitting
this paper).  We mostly avoided systems with known superhump 
periods, since their orbital periods can be deduced
fairly well from this information (see e.g. \citealt{unveils}).
JRT maintains a master list of CVs and updates it
regularly with new discoveries from the ASAS-SN, MASTER, 
and CSS lists, and other discoveries as they come to
his attention.  Before every observing run, this list
is searched and targets that might yield useful
spectra -- in practice, targets that are $\sim 19.5$ mag
or brighter at minimum -- are culled out.  Final 
selection is made at the telescope using quick-look
spectra and radial velocities measured in real time.
In practice, the selection process favored objects
with strong H$\alpha$ emission, and disfavored objects
with strong continua and weak emission such as dwarf
novae in outburst and some novalike variables. 

\section{Observations and Reductions}

All our observations are from MDM Observatory on
Kitt Peak, Arizona.  

{\it Spectroscopy.} For nearly all the spectra, we used the 
`modspec' spectrograph\footnote{A description of the modspec
can be found at 
{\tt http://mdm.kpno.noao.edu/Manuals/ModSpec/modspec\_man.html}.}
with a 600 line mm$^{-1}$ grating.
We mostly used a SITe 2048$^2$ CCD detector, which yielded 
2 \AA\ pixel$^{-1}$ from 4210 to 7500 \AA, with declining
throughput toward the ends of the spectral range.  When this 
detector was unavailable, we used a very similar 1024$^2$
SITe detector (`Templeton'), which covered from 4660 to 6730
\AA .  The modspec was mounted mostly on the 2.4 m Hiltner 
telescope, but for some of the brighter objects we used 
the 1.3 m McGraw-Hill telescope.  For a few of the 1.3-m
spectra we used the Mark III grism spectrograph, which 
covered 4580 to 6850 \AA\ at 2.3 \AA\ pixel$^{-1}$.  
On both telescopes, and with both spectrographs, we used
an Andor Ikon camera to view the reflective slit jaws
through a microscope, 
and guided the telescope with a separate off-axis guider.
With this arrangement we could place any object that was
bright enough for a usable spectrum in the slit, and 
track it accurately even if the portion of the light 
spilling onto the slit jaws was invisible.

Observations from a single site must be taken at intervals
of about one sidereal day, which inevitably leads to 
some uncertainty in the number of cycles that elapsed
when the object was out of view.  To minimize this uncertainy,
one must take observations far from the meridian. A
spectrograph slit that is oriented north-south -- the
default for these instruments -- will then have a position
angle far from the parallactic angle, along which 
atmospheric dispersion acts \citep{filippenko82}.  The
resultant wavelength-dependent decentering 
causes light loss that varies with wavelength, and
small spurious velocity shifts as well.  To avoid this,
at the 2.4m we rotated the instrument to align the 
1.1-arcsec slit with the parallactic angle when observing at large
hour angles.  At the 1.3 m, the effect is less critical because
of the larger projected slit width (2 arcsec) and the rotation
adjustment is much more awkward, so we did not rotate the instrument.

We derived the wavelength calibration from Hg, Ne, and Xe lamp 
spectra taken near the zenith in twilight. With modspec,
we did not take lamps when the sun was more than 12
degrees below the horizon, and instead used the 
$\lambda 5577$ night sky line to track and 
correct for drift in the zero point of the calibration.
Other night-sky features that were not used in the calibration 
typically had mean velocities $< 10$ km s$^{-1}$ and an 
RMS scatter of $\sim 3$ km s$^{-1}$.  The night-sky
technique did not give good results for the Mark III
spectrograph, so with that instrument we took lamp
spectra at each new telescope position, and at 
intervals of $\sim 1$ hr when following a target.

{\it Radial Velocities.}
Our emission-line radial velocities are almost entirely 
of H$\alpha$, since it almost always gives the best
signal-to-noise ratio with our instrument.  We use
an algorithm developed by \citet{sy80}, in which 
the line profile is convolved with an antisymmetric
function and the zero of the convolution -- which 
occurs when the positive contribution from one
side of the line balances the negative contribution
from the other side -- is taken as the
line center.  When measuring the doubled, steep-sided
lines of edge-on dwarf novae, we adjusted the convolution
function to be sensitive to the steep sides of the line.
For narrower emission lines the function we used was
the derivative of a gaussian, with the width tuned to the
line.  To calculate uncertainties in the velocities 
we propagated the estimated photon-statistics 
uncertainties of the spectrum points through the measurement
process; this will be optimistic, since it does not 
include any systematic effects (e.g., line-profile
variations).

When a late-type secondary star was present, we found
its velocity using the {\tt xcsao} task from the {\tt rvsao} 
package \citep{kurtzmink}, which implements the cross-correlation
velocity algorithm described by \citet{tonrydavis}.  
As a template, we used a composite of a large number 
of K-type spectral standards, shifted to zero velocity,
This matched the $\sim$ K-type secondaries reasonably well.
The algorithm estimates uncertainties using the asymmetry
in the cross-correlation.

To search for radial-velocity periods, we constructed a
grid of trial frequencies with spacing $\Delta f$ 
sufficiently dense for the 
time interval $T$ covered (in practice, $\Delta f = 1 / (12 T)$
for long time-span searches, and denser for shorter spans).
We then fit least-squares sinusoids of the form 
$v(t) = A \cos \omega t + B \sin \omega t + C$ at each trial
frequency $\omega = 2 \pi f$, and computed 
\begin{equation}
\chi^2 = \sum_i  {\left[o_i - v(t_i)\right]^2 \over \sigma_i^2}, 
\end{equation}
where $o_i$ is the $i$-th observed data point.  We 
display $1/\chi^2$, which is visually analogous to a 
periodogram, and which tends to emphasize frequencies
with very low $\chi^2$ (and hence very good fits).
To test the reliability of the choice of cycle count, we
used the Monte Carlo tests described by \citet{tf85},
and in nearly all cases continued taking measurements until the
likelihood of an incorrect choice was very small.   

All these techniques are incorporated in a quick-look 
reduction pipeline for use at the telescope. With this we 
monitor velocities in real time, update the period search,
and plan observations for maximum period discrimination.

{\it Photometry.} For nearly all our time-series differential
photometry, we used an Andor Ikon
camera (the same model used for slit viewing), 
attached to the 1.3 m McGraw-Hill telescope at MDM.  
All observations were taken either unfiltered, or  
with a GG420 filter to suppress moonlight. 
The 2015 March observations of ASAS-SN 15aa, and
the 2016 March observations of ASAS-SN 15cw,  were taken 
through the slit-viewing microscope, with the target
well away from the slit.  These images were cosmetically
poor but proved adequate for eclipse timing.  The rest of
the data were taken with the camera in direct imaging mode.  
Exposures were typically 10-30 sec, with almost no dead
time.  We subtracted bias and dark frames from 
the raw images, and divided by a flat field image taken
in twilight.  

To measure instrumental magnitudes we used
the IRAF {\tt phot} routine, with for the most part
a 3-arcsec radius aperture.  In
each image we measured the program star, a brighter
comparison star, and at least one check star.
Because of our non-standard passband we did not observe
photometric standard stars.  We instead used
catalog data to find or estimate $V$ magnitudes for the 
comparison stars, and added these to the differential magnitudes
to align them very roughly with $V$.

{\it Secondary star contributions and distances.}  
Several objects show a late-type
contribution due to a secondary star.  When this was
sufficiently distinct, we estimated its spectral type and
contribution in the following manner.  First, if the 
secondary's radial velocities were measurable, we 
shifted the individual spectra into the rest frame of
the secondary and co-added them.  We then scaled and subtracted
spectra of stars from a range of spectral type,
and adjusted the type and scale factor to best 
cancel the late-type features.  The 
spectral type and estimated flux contribution correlate
strongly, since metal features tend to grow stronger
toward cooler types.  

To estimate distances using the secondary contributions,
we used a Monte Carlo technique described in
\citet{thorhalpern13}.  Briefly, we use the orbital
period and a rough estimate of the secondary's mass to
constrain its radius (which depends only weakly on the
mass), and then use its spectral type to estimate its
surface brightness; combining these with its flux 
contribution then yields a distance.  Note that we do not
assume that the secondary follows
a main-sequence mass-radius-temperature relation.
The Monte Carlo technique picks all the input quantities from
assumed distributions, and in particular accounts
for the correlation between spectral type and 
secondary star contribution. 

\section{The Individual Stars}
\label{sec:individuals}

Table \ref{tab:star_info} lists the objects, 
their coordinates, and the shorter names we use here,
in order of right ascension.  

\begin{figure}
\includegraphics[height=23 cm,trim = 2.2cm 1cm 1cm 2.8cm,clip=true]{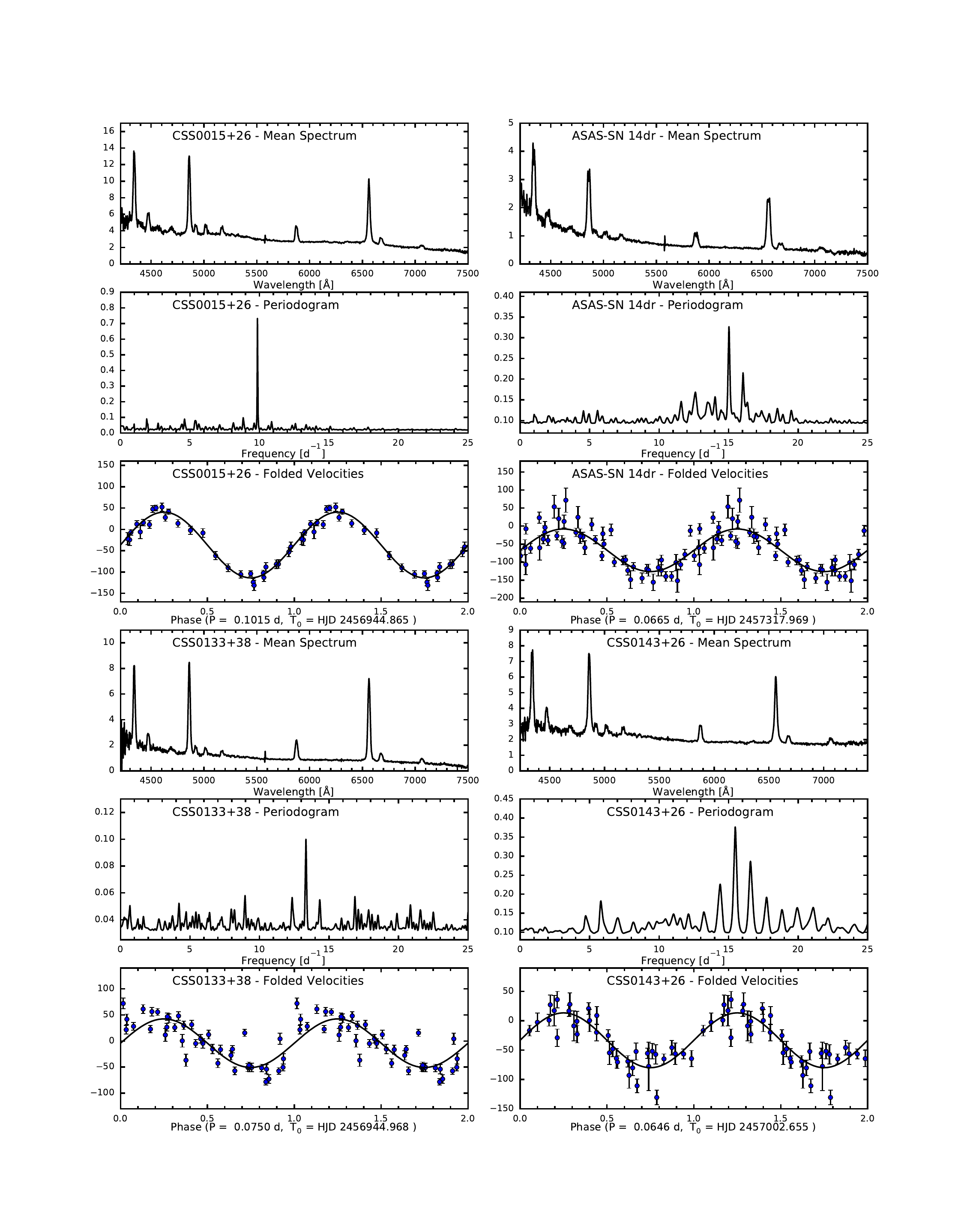}
\vspace{0.4 truein}
\caption{{\it Caption on next page.}}
\label{fig:cvplot1}
\end{figure}

Figures~\ref{fig:cvplot1} -- \ref{fig:cvplot9} show velocities,
velocity periodograms, and folded velocities for 33 of the objects
discussed here.  Table \ref{tab:velocities} lists all the radial 
velocities, and Table \ref{tab:parameters} gives parameters of 
sinusoidal fits to the velocities.  Table \ref{tab:eclipsetimings}
gives eclipse timings for the four objects in which 
we observed eclipses.

\subsection{CSS0015+26}

The CRTS alert on this source identified it with the
ROSAT X-ray source 1RXS J001538.2 +263656 \citep{drake14}.
The CRTS DR2 light curve shows a quiescent magnitude
near 18 mag for this dwarf nova, with three outbursts 
to $\sim 14$ mag between 2005 and 2013, the brightest outburst 
reaching 13.3.  \citet{szkody14} obtained eight spectra in
the blue, and suggested a period near 100 min.

Our mean spectrum  (Fig.~\ref{fig:cvplot1}) shows strong emission on a blue continuum,
similar to that published by \citet{szkody14} but covering 
a greater spectral range.
H$\alpha$ has an equivalent width (EW\footnote{We
quote emission-line equivalent widths as positive.})
of 80 \AA , and a full width at half-maximum (FWHM) of 
1200 km s$^{-1}$.  The He II emission line is present,
but much weaker than H$\beta$.  The synthesized $V$ is
17.7, consistent with minimum light.

We first observed this object in 2012 September, but
obtained only fragmentary data before the object went into
outburst on the night of 2012 Sep.~9 UT\footnote{The CRTS light 
curve missed this outburst.  The next CRTS measurement,
taken seven days later, showed the source at 17.1 mag.}.
We obtained more extensive data in quiescence
in 2014 October, which are best fit by
$P_{\rm orb} = 2.436(2)$ hr, toward the lower
end of the period gap but well inside it. 
Although dwarf novae in the gap are somewhat unusual, the
period is reasonably secure, since the data span over 7 hr of 
hour angle.  The adopted period results in much better
fits to the data than alternate choices of cycle count.

There seems to be no record in the literature of 
superhumps appearing in this star.  Vsnet-alert 
13818\footnote{Vsnet-alert is an email service to alert
observers to interesting targets; an explanatory 
page is found at 
{\tt http://ooruri.kusastro.kyoto-u.ac.jp/mailman/listinfo/vsnet-alert},
which links to monthly archives of past alerts.  Many archived
alerts were posted by T. Kato of Kyoto University. Archived 
alerts are challenging to reference in the traditional format,
since they are not indexed at ADS, SIMBAD, or other sites.  
We adopt the practice of citing these by number, and
where possible mentioning the observers who contributed.
We only cite vsnet when it provides information that we
have been unable to find elsewhere.}
describes the 2009 outburst as having ``quickly
faded'', and the 2012 outburst appears to have also
faded rapidly; a superoutburst may not have occurred.

\subsection{ASAS-SN 14dr} 


This dwarf nova was detected by ASAS-SN on 2014 July 8,
reaching $V \sim 15$.  In the CRTS light curve it is
typically $V \sim 18.5$ in quiescence, with 
variation by over a magnitude, and there are two
previous outbursts.  

The summed spectrum (Fig.~\ref{fig:cvplot1}) shows
a blue continuum with strong emission lines typical
of a dwarf nova at minimum light.  The synthesized
$V \sim 19.2$, and H$\alpha$ has a FWHM near 1800
km s$^{-1}$ and an emission EW of 140 \AA . 
The line is double-peaked, with the peaks separated
by $\sim 600$ km s$^{-1}$.   The velocity period,
95.83(13) min, is determined without significant
ambiguity.  At this period, we can 
anticipate the appearance of superhumps in a 
future outburst. 

\subsection{CSS0133+38}

\citet{drake14} identify this dwarf nova with the 
ROSAT X-ray source 1RXS J013308.9 +383218.
The average spectrum (Fig.~\ref{fig:cvplot1}) shows strong, single-peaked 
emission lines; the EW of H$\alpha$ is 220 \AA , and 
its FWHM is 1150 km s$^{-1}$.  The synthesized
$V$ magnitude is 18.9.  The CRTS DR2 light curve
shows a typical quiescent magnitude near 18.5, with 
occasional excursions as faint as nearly 20th magnitude,
and a single outburst up to 16.0 in late 2009; vsnet-alert
17667 reported a second outburst in 2014 August, but
it faded rapidly.  The radial
velocities indicate $P_{\rm orb} = 108.0(2)$
minutes, without significant ambiguity.  At 
this period, superhumps are to be expected if 
a superoutburst occurs.



\subsection{CSS0143+26}

This dwarf nova is apparently detected in the ROSAT Faint Source
Survey as 1RXS J014306.4 +263834.
The spectrum (Fig.~\ref{fig:cvplot1}) appears typical for a dwarf nova at minimum light;
H$\alpha$ has an EW of 75 \AA\ and a FWHM of almost
1200 km s$^{-1}$, and the $V$ magnitude synthesized from the spectrum
is 18.1.  The CRTS light curve shows a typical quiescent magnitude
near 17.5, with at least a half-magnitude of scatter around this,
and a single unambiguous outburst in 2006 December that reached
14.9 mag.  The radial velocities indicate $P_{\rm orb} = 
92.4(3)$ min, a range in which superhumps can be expected.



\subsection{OT0150+37}

\citet{lazavera13} discovered this CV in data from the MASTER
robotic telescope network, and identified it with the faint 
ROSAT X-ray source RXS J015017.0+375614.  The CRTS light
curve appears unusual, with irregular variations between
17th and 19th magnitude and occasional outbursts to brighter
than 15th.  Denisenko (2013; Vsnet-alert 16457) monitored the source
more closely and suggested that it might be an active polar,
or AM Her star.  \citet{szkody14} obtained spectroscopy on 
one night, indicating an orbital period near 103 min, which 
was corroborated by two humps seen in 2.8 hours of time series
photometry.  Because the HeII $\lambda 4686$ line was absent, and the 
velocity amplitude modest, they rejected the polar classification.

The mean MDM spectrum from 2013 October (Fig.~\ref{fig:cvplot2}) 
appears normal for
a dwarf nova, in agreement with the conclusions of \citet{szkody14}.
HeII $\lambda 4686$ does appear weakly, with an equivalent 
width of $\sim 5$ \AA. 
The source was in its intermediate state, with a synthesized $V = 16.7$.
The single-peaked H$\alpha$ emission line had an EW of 60 \AA\ and a FWHM
near 1000 km s$^{-1}$.  The radial velocities clearly indicate
$P_{\rm orb} = 213.3(4$) min, or 3.555(7) hr.  Because the present data are much
more extensive than those of \citet{szkody14}, and have a 
much greater hour-angle span, this period supersedes their 
103-min period.   The two bumps seen in their photometry may arise
from random variation rather than representing a sustained
periodic modulation; more extensive time-series photometry
would be needed to decide this.

The previous month we had obtained two spectra on
2013 Sept.~11 UT, and a single spectrum on Sept.~15.  
The latter showed the source in outburst, with weak
Balmer emission on a strong blue continuum, and a synthesized
$V$ magnitude of 15.4.  None of our spectra were taken
in the 19th-magnitude deep low state. 

The outburst spectrum and weakness of HeII $\lambda 4686$
both support a dwarf nova classification.  However, 
the light curve renders the classification somewhat tentative.


\subsection{CSS0208+37}

This CV was discovered on 2010 December 7, and lies
28 arcsec from 1RXS J020802.6+373236.  \citet{thorskinner}
obtained a spectrum, but no time-series observations.
The CRTS light curve shows a fairly steady quiescent 
magnitude near 17.8, with outbursts in 2006, 2008, and 2010.

Vsnet-alert 12467 reports detection of a 0.45 mag oscillation with a
a period of only 0.0193 d, or 27.8 min, and credits the
detection to Maksim Andreev of the Terskol team.
If such a short period were orbital it would suggest a
double-degenerate CV, or AM CVn star, but the spectrum
(Fig.~\ref{fig:cvplot2}),
which appears similar to that shown by \citet{thorskinner},
but with better signal-to-noise,
has normal Balmer emission lines and
resembles the spectrum of an ordinary dwarf nova.
H$\alpha$ has an EW of $\sim 80$ \AA , a FWHM of
2000 km s$^{-1}$, and a double-peaked profile with
peak separation $\sim 825$ km s$^{-1}$.  

We have observations from several observing runs, 
the most extensive being 2013 January.  The H$\alpha$
radial velocities define an unambiguous period near
84.2(2) min, consistent with expectations from the
spectrum.  The spectroscopy and the light curve are
both consistent with a normal short-period dwarf nova. 
We cannot shed any light on the 27.8-min photometric 
oscillation.

\subsection{ASAS-SN 14dx}

This transient was detected by ASAS-SN on 2014 July 12;
they noted that it corresponds to a GALEX and XMM-Newton
source. \citet{kaur14} obtained spectra in late August;
their description 
is consistent with the 2014 October spectra presented here
(Fig.~\ref{fig:cvplot2}).
The CRTS DR2 data show the target as consistently
near 16.2 mag, with one detection at 16.6.  
In vsnet-alert 18017, T. Kato reports that Josch Hambsch 
and Berto Monard detect a persistent photometric
period of 0.05750 d; Kato was able to tie this to 
the long-term CRTS light curve and refine the photometric
period to 0.0575060(2) d, or 82.8086(3) min.

The mean spectrum shows a strong, blue continuum with 
double-peaked emission lines.  The H$\beta$ and H$\gamma$ 
lines (the latter near the end of the spectral range) are
flanked by absorption, very likely contributed by the 
white dwarf.  The continuum appears `wavy', but the 
`waviness' is unlikely to be real, because the modular
spectrograph is known to sometimes produce irreproducible waves
in continua, which we are unable to calibrate out.  
The EW of H$\alpha$ is 95 \AA , its FWHM is 1400 km s$^{-1}$,
and its double peaks are separated by a bit more than
500 km s$^{-1}$.  The H$\alpha$ radial velocities 
are modulated at an unambiguously-defined period
of 82.88(8) min, consistent with the 
more accurate photometric period, and confirming that
the photometric signal is orbital in nature.

It is unusual to discover a new dwarf nova 
with such a bright minimum magnitude.  The URAT1 
astrometric catalog \citep{urat} gives a proper motion 
$[\mu_X, \mu_Y] = [+83(7), -37(7)]$ mas yr$^{-1}$, 
derived from the difference between the URAT1 and 2MASS positions. 
At 100 pc, this corresponds to a transverse velocity 
of 43 km s$^{-1}$, which suggests a distance less than
a few hundred parsecs.
The star is listed in SDSS DR12 with $g = 16.31$,
$u-g = 0.13$, $g - i = -0.26$, $r - i = -0.200$, and
$i - z = -0.058$.
 
%

\subsection{CSS0357+10}

\citet{schwope12} reported photometry and archival X-ray observations
of this source, and suggested it as a candidate polar, or AM Her
star.  They used archival CSS photometry to find a photometric
period of 0.079181(1) d, or 114.021(1) min.  

Our spectra (Fig.~\ref{fig:cvplot2}) 
were taken with the smaller `Templeton' detector,
so the HeII $\lambda 4686$ line is near the end of our
coverage, but it is nearly as strong as H$\beta$;
such strong $\lambda 4686$ is characteristic of 
magnetic systems.  H$\alpha$ has an EW of 31 \AA\ 
and a FWHM of 900 km s$^{-1}$.  The radial velocities
are modulated with a large semi-amplitude $K = 320(19)$ 
km s$^{-1}$ at $P = 0.07918(6)$ d, identical to
the \citet{schwope12} period, but less precise due
to the shorter time base.  The strong HeII and large
velocity amplitude strongly corroborate the
suggestion by \citet{schwope12} that this is a polar.

\subsection{CSS0500+13} 

S. Kaneko discovered this object at magnitude 12.3 on 
2015 November 28 (vsnet-alert 19307), 
and shortly afterward CSS found it independently.  Its
CRTS light curve shows irregular variations from roughly
16 to 17 magnitude, with one outburst to 14.3 in 
2007 December, and another measurement at 15.4 in 
2008 November; outbursts are evidently rather infrequent.
The CV is the likely optical counterpart
of the X-ray source 1RXS J050027.7+133415.  Kato
(vsnet-alert 19312) noted that the SDSS colors 
($g = 18.59, u-g = 0.63, g - 4 = 1.03, r - i = 1.03, 
i-z = 0.78$) are unusually red for a CV.  The
SDSS magnitudes are significantly fainter than expected
given the CRTS light curve, suggesting the source was  
in an atypically low state when the SDSS direct images were
taken.

The spectrum
(Fig.~\ref{fig:cvplot3}) shows a strong contribution
from a secondary star of type M3.5 $\pm 0.5$. 
The SDSS $r-i$ color is similar to an
M2 dwarf, while $i-z$ is closer to expectations for
M4 \citep{west08},  which suggests
that the shorter passband included some
accretion light. The spectrum has a
synthesized $V = 17.7$, whereas 
R. H. Lupton's transformation 
\footnote{See {\tt http://www.sdss.org/dr7/algorithms/sdssUBVRITransform.html}} 
from SDSS $g$ and $r$,
namely $V = g - 0.5784 (g - r) - 0.0038$,
yields $V = 17.99$.  H$\alpha$ is 
strongly in emission with an EW of 
$\sim$ 130 \AA\ and FWHM of $\sim$1600 km s$^{-1}$. 
The Balmer lines are double-peaked, with a
separation of $\sim 740$ km s$^{-1}$. 

The H$\alpha$ line yields a remarkably clean
modulation at $P = 3.7116(6)$ hr.  
We observed the object on closely-spaced observing
runs in 2016 February and March, and there is no
ambiguity in the cycle count over that interval. 
The secondary's spectral type is approximately as
expected at this $P_{\rm orb}$ \citep{kniggedonors}.
From the secondary star's contribution (see Section 2)
we estimate the distance to be $340 \pm 80$ pc.

\subsection{OT0647+49}

This MASTER transient was discovered in images taken 2013 March 7
\citep{tiurina13}.
It proved to be an SU UMa-type dwarf nova; \citet{kato14} 
found superhumps at 0.067774(41) d, evolving to 0.067337(31) d.  
The H${\alpha}$ emission line (Fig.~\ref{fig:cvplot3}) is broad, 
with a FWHM of 2100 km s$^{-1}$,
and double-peaked, with the peaks separated by $\sim 940$ km s$^{-1}$,
indicating that the orbital inclination is not far from edge-on;
its EW is $\sim 60$ \AA .

The radial velocities select a period near 0.0655 d, or 94 min.
Nearly all our data are from two runs, in 2014 January and
2014 December.  Separate fits to the two runs' data yielded
the same period within the errors, and the weighted average
(quoted in Table \ref{tab:parameters}) amounts to 94.4(2) min.
Taking superhump period $P_1$ from \citet{kato14} and computing
the superhump period excess, 
$$ \epsilon = {P_{\rm sh} - P_{\rm orb} \over P_{\rm orb}},$$
gives $\epsilon = 0.034$.  \citet{unveils} give a relation
to predict the orbital period from the superhump period, which 
for the two periods quoted by \citet{kato14} gives
0.0660(5) and 0.0656(5) d, respectively; both predictions
agree reasonably well with our measured period.  This appears
to be a normal SU UMa star.

\subsection{ASAS-SN 14kj} 


Little is known about this object.
The CRTS light curve is sparse, showing 
variation from 17 to 19 mag and no detected outbursts.
ASAS-SN detected it at 14.9 mag on 2014 Nov. 17.
SDSS DR12 lists it at $g = 18.44$, and our mean spectrum
(Fig.~\ref{fig:cvplot3})
has a synthesized $V = 18.66$.  The spectrum appears
normal for a dwarf nova at minimum light;  H$\alpha$
has an EW of $\sim 80$ \AA\ and a FWHM of 1500 km s$^{-1}$.
The period, 130.4(4) min, is determined without
significant ambiguity because the observations span 
7.9 h of hour angle; this period lies near the lower boundary
of the roughly 2-3 hr gap in the CV period distribution.  
Superhumps may be expected in a future outburst.

\subsection{ASAS-SN 15cw} 


This object was discovered by ASAS-SN on 2015 Feb.~11, and reached
a magnitude of 15.9.  The CRTS DR2 shows it varying irregularly
between about 17th and 20th magnitude, but does not show any
outbursts.  Our mean spectrum (Fig.~\ref{fig:cvplot3}) 
has synthesized $V = 18.4$. 

We took spectra in 2016 February, and quickly found 
signs of an eclipse, in particular a strong rotational
disturbance (or Rossiter-McLaughlin effect; \citealt{rossiter24},
\citealt{mclaughlin24}) in the H$\alpha$ emission line -- the line
center shifts abruptly to the red as the leading (blueshifted) 
side of the disk is eclipsed, and then to the blue as the 
leading side emerges, leaving the redshifted trailing side 
hidden (see Fig.~\ref{fig:cvplot3}).  Time-series photometry on the same run
showed the eclipses to be at least two magnitudes
deep and recurring with a period near 113.9 min
(Fig.~\ref{fig:15cweclipse}).
The H$\alpha$ radial velocities away from eclipse
were consistent with this period, but in the interest 
of observing efficiency
we did not obtain the extensive spectroscopy 
needed to establish $P_{\rm orb}$ 
from the velocities alone.  We obtained a final
eclipse timing 2016 March with the slit-viewing camera
on the 1.3 m telescope; including this yields
for the Barycentric Julian Date at mid-eclipse
$$
\hbox{BJD mid} = 
2457431.7674(1) +  0.0791258(4) E,
$$
where $E$ is the cycle count.  We attempted to refine
this using the CRTS DR2, but (oddly) could not discern the 
eclipse in that data set.

Although the system must be nearly edge-on, the
emission lines (Fig.~\ref{fig:cvplot3}) are not notably
double-peaked.  H$\alpha$ has an EW of 130 \AA\ and a
FWHM of 1600 km s$^{-1}$, so they are at least rather
broad, as expected. 

\subsection{ASAS-SN 14ag} 

ASAS-SN discovered this source on 2014 March 14 at 13.5 mag.
The CSS light curve shows several outbursts to nearly this 
brightness from a typical quiescent magnitude just fainter
than 16, as well as a few points near 17th.  \citet{katovii}
found eclipses of $\sim 1.5$ mag depth in the CRTS data, derived an eclipse
ephemeris with a period near 86.9 min,
and also found superhumps with $P_{\rm sh} = 0.062059(55)$ d,
or 89.37(8) min.

We have spectra from 2014 December and 2016 January.
The mean spectrum from 2016 January (Fig.~\ref{fig:cvplot4}) 
is typical for a 
short-period, nearly edge-on dwarf nova near minimum light.
The H$\alpha$ line has an EW of 90 \AA , a FWHM of
1900 km s$^{-1}$, and double peaks separated by about 
790 km s$^{-1}$.  The double-peaked nature of the 
HeI lines is more pronounced, with the line centers 
extending down to the continuum, or, in $\lambda\lambda$ 
4471, 4921, and 5015, even below the continuum.  
In 2016 January we also obtained time series photometry
(not shown) that yielded five eclipses.

The H$\alpha$ radial velocities from 2016 January 
independently constrain $P_{\rm orb}$ to 0.06029(5) d; 
including the 2014 December data yields multiple precise
periods separated by 1 cycle per 397 d.  One of these
precise periods agrees accurately with Kato's candidate
eclipse period from the CRTS; Kato's ephemeris also 
predicts our 2016 January eclipses accurately, confirming
his choice of period.  Adopting his period and using
our 2016 January eclipses to set the epoch yields,
for mid-eclipse,
$$\hbox{BJD mid} = 2457416.7385(1) + 0.060310651(2) E
$$
where $E$ is an integer cycle count.  
We use this for 
folding the velocity data in Fig.~\ref{fig:cvplot3}.
Note that the radial velocities do not decrease through their
mean at eclipse phase, as would be expected if they
traced the motion of the motion of the white dwarf. 

Like ASAS-SN 14dr, this system is unusually bright for a
newly-discovered dwarf nova, suggesting it is relatively
nearby.  The URAT catalog \citep{urat}
gives its proper motion as $(\mu_X,\mu_Y) = (-54, +72)$ mas
yr$^{-1}$, with an error of about 5 mas yr$^{-1}$ in each 
coordinate.  Using this information alone, and assuming a
velocity distribution typical of disk stars, the 
procedure described by \citet{thorparallax03} gives
a likely distance of only $\sim 60 (+40,-21)$ pc. 
\citet{warn87} developed a method of estimating 
the absolute visual magnitude of a dwarf nova at maximum 
light. While this involves a correction for inclination, the
eclipse constrains the inclination of this system to be
close to edge-on.  If we take the inclination to be $\sim 80$ 
degrees, and use $V=13.5$ for maximum brightness,
Warner's method gives a best distance of $\sim 90$ pc, also
quite nearby for a newly-discovered system.  

\subsection{ASAS-SN 14ak} 


This transient was detected by ASAS-SN on 2014 April 5 at
15.4 mag.  It appears to be the optical counterpart of
1RXS J095056.2$-$294649.
The CRTS DR2 light curve is fairly well-sampled, with 
260 points over eight seasons.  It shows the source varying by
a half-magnitude or so around a mean magnitude of $\sim 17.6$, 
with one outburst in 2009 April that reached 14.15 and another
in 2010 May that reached 14.9.  The mean spectrum 
(Fig.~\ref{fig:cvplot4}) has a synthesized
$V = 18.4$.  H$\alpha$ has a FWHM $\sim 1300$ km s$^{-1}$ 
and an EW of $\sim 130$ \AA ; its profile is not double-peaked,
but the HeI lines show two peaks at $\pm 400$ km s$^{-1}$ from the 
line center. 

Because of the southerly declination, our data span only
5.5 hr of hour angle.  Nonetheless, the period is determined without
significant cycle count ambiguity as 127.7(1) min (11.28 cycle d$^{-1}$), 
with a daily alias at 140.7 min (10.24 cycle d$^{-1}$) being much less 
likely.  This places the system near the lower edge of the period
gap, and superhumps may be expected in the event of a superoutburst.

\subsection{CSS1028$-$16}


This object is apparently the optical counterpart of 1RXS J102844.2$-$161253.
Its CRTS DR2 light curve is somewhat unusual -- it
mostly stays near 17th magnitude but has many excursions to
nearly 20th, and no obvious outbursts.  The downward 
excursions are persistent and do not appear to indicate
eclipses.  Despite the odd variability, the
mean spectrum {Fig.~\ref{fig:cvplot4})} 
appears ordinary for a CV.  The single-peaked
H$\alpha$ line has an EW of 75 \AA\ and FWHM $\sim 1300$ km
s$^{-1}$, again not suggesting an especially high inclination.  
Our spectrum has a synthesized $V \sim 18.3$, near the middle 
of the range of variability.  There is a contribution from an M-type
secondary.  While this is too weak to usefully constrain the secondary's
spectral type or the system's distance, it does suggest that the 
luminosity was not especially high during our observations.  
The velocity data span 7.0 hr of hour angle, and determine the 
period uniquely as 4.553(14) hr.  

The unusual variability suggests a magnetic system, 
but HeII $\lambda4686$ does not appear in our mean spectrum, 
so CSS1028$-$16 is likely to be some other kind of CV, possibly a Z Cam 
star (a dwarf nova with bright `standstills'). 
 
\subsection{SDSS 1029+48}


This object was discovered by \citet{kepler16} in a search for
new white dwarfs in the SDSS DR12.  It lies 3.5 arcsec north
of a slightly brighter M-type star.  The CRTS light curve,
which probably includes both stars, is steady around magnitude
16.15 with a single possible brightening in 2006 to $\sim 15.85$.
so outbursts are infrequent or possibly absent.

Our spectrum (Fig.~\ref{fig:cvplot4})
resembles the SDSS spectrum, showing emission lines with double
peaks $\pm 500$ km s$^{-1}$ from the line center. 
H$\beta$ shows weak absorption flanking the emission, which
probably arises from the white dwarf photosphere.
The blue continuum also suggests that the
white dwarf contribution is strong.  Some M-type
features appear weakly toward the red, but these are
almost certainly contamination from the neighboring M star,
which we were unable to exclude completely. 

We have spectra from two runs about 16 days apart, in 2016 
January and February.  The radial velocity modulation is weak, 
but indicates $P_{\rm orb} = 91.33(1)$ min. An alternate, but
unlikely, choice of cycle count between the runs gives 
91.11 min, and a still more unlikely choice of daily cycle count 
gives 85.87 min.

On 2016 June 11 we obtained 92 min -- just one orbital period
-- of continuous unfiltered differential 
photometry with the Andor camera on the 1.3 m telescope.
On all but a few of the 30 s exposures the seeing was good 
enough to split the target reasonably well from its neighbor.  
Excluding a few points taken in poor seeing,
the RMS scatter of the points was 0.04 mag.  
The light curve was featureless at this signal-to-noise,
and in particular showed no eclipses.

Fig.~\ref{fig:trails} shows a phase-averaged greyscale representation of our
spectra in the region of HeI $\lambda 5876$.  A prominent 
$S$-wave is present, which is formed where the mass-transfer
stream strikes the disk.  A similar $S$-wave is discernible
in H$\alpha$, but is less conspicuous because of the strong 
double-peaked emission line.  Tracing the HeI $S$-wave by eye,
and fitting it with a sinusoid, gave a semiamplitude $K_S \sim 540$
km s$^{-1}$ and phase lagging by about 60 degrees from the 
fit given in Table \ref{tab:parameters}. 

Emission line radial velocity amplitudes are 
notoriously difficult to interpret.  The velocity
amplitude $K = 22(4)$ km s$^{-1}$ listed in Table 
\ref{tab:parameters} is probably confounded by the 
$S$-wave contribution, and should not be taken 
to indicate the white dwarf's projected velocity
amplitude $K_1$.  However, measurements of the 
`horns' of the H$\alpha$ line profile suggest that 
$K_1$ is indeed rather low, probably 
less than $\sim 30$ km s$^{-1}$.  Adopting this as an
upper limit would imply $K_S / K_1 > 18.$  We can use this 
to set an inclination-independent upper limit on the mass ratio $q = M_2/M_1$,
guided by integrations of particle trajectories 
in the Roche potential, as follows.  The $S$-wave comes from the 
impact of the mass-transfer stream on the disk. 
The disk radius is unknown, but it must be
at least as large as the periastron distance of 
the mass-transfer stream, and it is likely to be 
significantly larger.  As a rough upper limit to the $S$-wave's 
in-plane velocity, we take the Keplerian speed at 
the periastron the gas stream would reach if it were not
to collide with the disk.
The hot spot's velocity is likely to be greater than the 
local Keplerian speed, because it includes gas from the fast-moving
stream, but the hot spot is likely to be at a significantly larger
radius than the periastron, so this should be 
a reasonable upper limit.  Using a 
nominal $M_1 = 0.7$ M$_{\odot}$, we find $K_S / K_1 = 18$
for $q \sim 0.15$, so that $q < \sim 0.15$.  
This is not unexpected -- \citet{pattlate} gives
$q \sim 0.1$ for stars near this orbital period.  

Since this is an upper limit, it is possible 
that the mass ratio is much lower, which would be 
significant.  Late in their evolution, CVs 
with hydrogen-rich secondaries evolve toward
a minimum period near 75 min as their 
secondaries lose mass; in theory, further
mass transfer should increase the secondary's
radius, and increase the orbital period.  
CVs that have evolved past the period minimum
are known as `post-bounce' systems.  They should
have periods somewhat longer than the minimum and
anomalously low mass ratios 
(\citealt{zharikov13} 
give a recent tabulation of post-bounce candidates).
At 91 minutes, with an evidently low mass-transfer 
rate and a possible low mass ratio, this system may 
be post-bounce.

\subsection{ASAS-SN 15aa}


This first entry in the ASAS-SN list for 2015 was detected on 2015 Jan.~1 
around 13.8 mag, and was matched to a $g = 16.8$ quiescent source.  The 
CRTS light curve, which extends from late 2005 to mid-2013, has 
289 detections over eight seasons, and detects three previous outbursts, 
one of which reached 13.8 mag on 2006 July 22.

The mean spectrum from 2015 March (Fig.~\ref{fig:cvplot5})
shows a strong contribution from a
late-type star.  We classify this as K4.5 ($\pm 1$ subclass)
and estimate that it contributes about half the light in the $V$ band.  
The fluxed spectra imply $V \sim 17.4 \pm 0.3$ for the secondary
contribution alone, where the uncertainty reflects only the 
range of acceptable spectral decompositions.  This estimate is
probably on the faint side because of occasional cloud and losses at
the spectrograph slit, so we adopt $V = 17.2$ for the secondary.

The radial velocities of the late-type star are modulated with a period
near 0.3755 d, or 9.01 hr, and a sizable semiamplitude $K_{\rm abs} = 
175(9)$ km s$^{-1}$.  The emission-line velocities show greater
scatter, but corroborate this period.  The spectral type of the 
donor star is normal for this period; it is cooler than 
predicted by a main-sequence mass-radius relation, but 
similar to the secondaries of other CVs in this period 
range \citep{kniggedonors}.

The CSS magnitudes (excluding outbursts) show a variation with two
maxima per orbit, and unequal minima, which are clearly ellipsoidal.
We used these data to refine the ephemeris as follows.  First,
we fixed the epoch to barycentric JD (BJD) 2457109.974(4), which is 
the blue-to-red crossing of the
absorption line velocities.  This corresponds to inferior conjunction
of the secondary (i.e., the phase at which an eclipse of the white
dwarf would occur for an edge-on system), and its  
uncertainty amounts to $\sim 0.01$ cycle.  
Then, we folded
the CSS magnitudes until the deeper minimum was both reasonably
sharp and coincided with the chosen epoch; Fig.~\ref{fig:crtsphot} shows
the result.  Because the CSS data
extend nearly 10 years earlier than the chosen epoch, the period
is greatly refined to 0.375540(2) d. 

We obtained most of our spectra in 2015 March, using the 
1.3 m telescope remotely.  The absorption-line orbit and its
large amplitude were immediately apparent, and suggested
the possibility of an eclipse, so we obtained time-series
photometry covering a predicted eclipse interval on 2015 March 26 UT, 
using the slit-viewing camera in unfiltered light, as 
described earlier.  We were 
rewarded by the discovery of a shallow eclipse.  Another eclipse light
curve was obtained three days later, and on April 10 we obtained
a third light curve, again remotely, using the Andor camera
in direct mode, again without a filter.  Two further eclipse light 
curves were obtained in 2016 January.

Fig.~\ref{fig:15aaeclipse} shows the light curves.  
The eclipse is very shallow, $\sim 0.25$ mag on the
ingress side, and is asymmetric, recovering only by 0.1 to 0.2 mag. 
in egress.  The center of the eclipse occurs around phase 0.007
in the absorption line ephemeris, which is consistent with
zero.  Because the phase of eclipse center corresponds accurately
with the inferior conjunction of the red star, it seems likely
that the white dwarf and the brightest regions at the disk center 
are being eclipsed.  The asymmetry may be due to the
occultation of the `hot spot' where the mass transfer stream
strikes the outer edge of the accretion disk.  
For our final ephemeris, we adjusted the epoch slightly
to correspond to the midpoint between the steepest portions
of the ingress and the egress in the April 10 light curve, 
yielding for Barycentric Julian Date of the inferior 
conjunction of secondary
$$\hbox{Inf. conj.} = \hbox{BJD } 2457109.977(3) + 0.375540(2) E.$$
This was used to compute the phases in Figs.~\ref{fig:cvplot5} 
and \ref{fig:15aaeclipse}. 


The emission and absorption velocity curves are perfectly antiphased,
to within the uncertainty.  This is a necessary, but not 
sufficient, condition for the emission lines to trace the motion
of the white dwarf.  The H$\alpha$ line profile appears
steady around the orbit, with two roughly equal peaks 
separated by $\sim 750$ km s$^{-1}$, shifting smoothly with 
the $K$-velocity, as would be expected from a symmetrical
accretion disk.   While the emission line velocities are
not nearly accurate enough to define a precise mass ratio,
their best-fit amplitude is a little less than the absorption
amplitude, suggesting a mass ratio not too far below unity.
The masses appear to be broadly as expected; the amplitudes 
are consistent with a 0.6 M$_\odot$ white dwarf and a 
0.55 M$_\odot$ red star in a nearly edge-on orbit (required
by the eclipse). 

Using the secondary star's contribution and the 
Monte Carlo procedure described in Section 2, we estimate 
the distance to be $1140 \pm 250$ pc.  The system
is very unlikely to be closer than 500 pc or more distant
than 2 kpc.  

\subsection{CSS1055+09} 


The CRTS light curve of this early CSS discovery shows
irregular variation between magnitudes 17.0 and 19.5 mag, 
with occasional outbursts to near 14th. 
It was discovered also in the
SDSS;  \citet{szkodyviii}
published a spectrum in quiescence showing emission lines with 
a sharp peak toward the blue, and suggested this might 
be from an S-wave that happened to be blueshifted at the time
of exposure.
\citet{thorskinner} also published a quiescent spectrum, and
noted a contribution from an M3$\pm$2 secondary. 

Fig.~\ref{fig:cvplot5} shows the mean spectrum from the 
present data.  The M-dwarf is again present but is less
prominent, probably because we caught the object in a brighter
state; the synthesized $V$ is 17.6, toward the bright
end of its quiescent range.  The emission lines are 
well-defined, the EW of H$\alpha$ being 70 \AA .
The H$\alpha$ velocities give an unambiguous 
$P_{\rm orb} = 3.90(0.01)$ hr.  A phase-averaged
greyscale representation of the H$\alpha$ line
(Fig.~\ref{fig:trails}) shows the orbital modulation very 
clearly, and reveals a slight doubling of the line.
The inclination is evidently rather high; eclipses
are not evident in Fig.~\ref{fig:trails}, but time
series photometry might reveal one.

Using both both the present mean spectrum and the data
from \citet{thorskinner}, we refined the secondary's
spectral type to M$2.75 \pm 1.25$.  The secondary-based
distance estimate (Section 2) yielded 730 (+200, $-$160) pc.  

\subsection{CSS1211-08} 

The CRTS light curve of this object shows it fairly 
steady between 17.5 and 18.5 magnitude, except for a
single outburst to 15.2 mag in 2011 February.
\citet{thorskinner} published a spectrum showing
strong Balmer and HeI lines.  Our mean spectrum
(Fig.~\ref{fig:cvplot5}) appears similar, but with
better spectral coverage and signal to
noise.  The Balmer lines are single-peaked;
H$\alpha$ has an equivalent width of 70 \AA\
and a FWHM of about 780 km s$^{-1}$, indicating
a low orbital inclination. 

The emission-line velocities from 2016 February span
almost 6.8 hours of hour angle, so the daily cycle
count is reasonably secure, giving $P_{\rm orb} = 113.9(3)$ min.  
At this period, superoutbursts and superhumps may
be expected.

\subsection{SDSS 1429+00} 

This relatively bright CV was discovered by \citet{carter13} in a search
for AM CVn-type binaries.  The CRTS DR2 light curve shows
it varying irregularly between 15.0 and 15.7 mag, with 
no outbursts. 
Our spectrum (Fig.~\ref{fig:cvplot5})
shows a strong blue continuum which approximately 
follows a power law, $f_\lambda \propto \lambda^{-2.4}$, and
relatively weak emission lines (the EW of H$\alpha$ is $\sim 22$ \AA ).
HeII $\lambda$4686 emission has an EW $\sim 3$ \AA, and CIII/NIII 
blend near $\lambda4640$ is comparable in strength. 
The spectrum and lack of outbursts are consistent with a 
novalike variable.

The radial velocities give an unambiguous period of 
3.560(4) d.  Fig.~\ref{fig:sdss1429trail} shows phase-averaged 
greyscale representations of the H$\alpha$ 
emission. The most remarkable feature is a weak, broad, high-amplitude $S$-wave
component with a semiamplitude of $\sim 1000$ km s$^{-1}$.
The line core appears to be double-peaked, with the 
peaks at $\pm \sim 280$ km s$^{-1}$ from the line center.
The high-amplitude feature is reminiscent of a still
higher-amplitude $S$-wave found in V795 Herculis
by \citet{haswell94}.

\subsection{ASAS-SN 15dl} 

This was discovered by ASAS-SN on 2015 February 16, at 15.6
magnitude.  The CRTS light curve shows at least 7 outbursts since
coverage began in 2005, reaching as bright as 14.2.  In 
quiescence, the object ranges from $\sim 18$ to $\sim 16.5$ mag.

We obtained a few spectra in 2015 March with the 1.3 m and 
modspec, but found the source in outburst and the H$\alpha$
line correspondingly weak.  In 2015 April the object
had returned to quiescence, so we obtained more
spectra with the 1.3 m and Mark III.  The average
quiescent spectrum (Fig.~\ref{fig:cvplot6})
shows Balmer and HeI emission, but
no HeII.  H$\alpha$ is single-peaked, with an equivalent
width of 68 \AA\ and a FWHM of 740 km s$^{-1}$.  The 
synthesized $V$ magnitude is 18.6, rather fainter than the 
CRTS light curve.

The radial velocities are fitted best with a period near
5.49 hr, but a daily cycle-count alias near 7.14 hours 
remains possible; the data span a relatively meager 5.1 hr of 
hour angle because of the object's southerly declination and 
the time of year.  The Monte Carlo test gives a discriminatory 
power of 0.95 for the alias choice, and a correctness 
likelihood of better than 99 per cent.  The choice of 
the shorter-period alias is also corroborated by absence
of secondary-star absorption features in the average spectrum.

\subsection{CSS1631+10}

This system is a known SU UMa-type dwarf nova, 
for which \citet{kato_ii} give
$P_{\rm sh} = 0.063945(24)$ d, or 92.1 min.  The CRTS DR2 light curve
shows significant variation in quiescence, between $\sim 17$ and
$\sim 19.5$ mag, and at least 5 outbursts between early 2005
and late 2013.  

The mean spectrum (Fig.~\ref{fig:cvplot6}) 
shows broad emission lines with incipient
double peaks; the EW of H$\alpha$ is 95 \AA\ and the FWHM is
around 2000 km s$^{-1}$.  The inclination is evidently not
far from edge-on, but eclipses have not been reported.  Since
superhumps are known, it is likely that deep eclipses would
not have been missed. 

The emission-line radial velocities are modulated at a period
near 90.2 min, just as expected from the superhump 
period, for which the mean $P_{\rm sh}$-$P_{\rm orb}$ relation
in \citet{unveils} would predict $90.0 \pm 0.7$ min.
While the daily cycle count is secure, we
have velocities from two observing runs $\sim 44$ d
apart, and the cycle count is ambiguous over that
interval.  Precise periods based on allowed choices
of the run-to-run cycle count (which give periods
within 3 standard deviations of the periods derived
from the individual runs) obey
$$P_{\rm orb} = {44.723(4) \hbox{ d} \over 714 \pm 4},$$
where the denominator is an integer.

The spectrum, and the as-expected relation between
$P_{\rm orb}$ and $P_{\rm sh}$, show this to be a typical
SU UMa-class dwarf nova.

\subsection{CSS1702+55}

This object was detected by CRTS on 2013 April 16.  Its
CRTS DR2 light curve shows variations at minimum from
18.0 to nearly 20th mag, and a single outburst to 
16th magnitude corresponding
to its discovery.  It is apparently the optical counterpart
of the ROSAT X-ray source 1RXS J170207.1+551746.

Our mean spectrum (Fig.~\ref{fig:cvplot6}) 
shows very strong, relatively narrow
single-peaked emission lines; the EW of H$\alpha$ is 180 \AA , and its
FWHM is 830 km s$^{-1}$.  In part because of the strength
of the lines, the radial velocities show relatively little
scatter around the best-fit sinusoid at $P_{\rm orb} = 99.8(2)$ min.
The radial velocity amplitude $K = 35 \pm 5$ is relatively 
low, which together with the relatively narrow lines suggests
that the orbital inclination is fairly low.  
At this period, superoutbursts and superhumps are expected,
but none appear to have been reported as of yet.


\subsection{OT 1727+38} 

\citet{denisenko14} detected this source in outburst at 
$V = 14.3$ using MASTER, and
noted previous outbursts detected on sky survey plates and in SDSS.
The CRTS DR2 light curve shows it mostly at a rather steady
minimum near 17.8 mag, with at least nine outbursts
between 2005 and 2013, reaching as bright at 14.6 mag.

Our mean spectrum (Fig.~\ref{fig:cvplot6}) has a synthesized
$V \sim 18.5$, and appears to have been taken in quiescence.
It shows typical CV emission lines on a blue continuum,
with weak absorption flanking the H$\beta$ emission,
suggesting a contribution from a white dwarf photosphere;
the steadiness of the minimum magnitude also suggests a 
stellar contribution.  
The H$\alpha$ emission line has a relatively modest equivalent 
width of 43 \AA , and is slightly double-peaked, with red and 
blue peaks at $~ \pm 320$ km s$^{-1}$ from the line 
center.  The FWHM of the line is $\sim 1300$ km s$^{-1}$.

The emission-line velocities indicate $P_{\rm orb} = 82.14(6)$ min.
This is the shortest period in the present sample.
The choice of daily cycle count is reasonably secure;
the Monte Carlo test gives a discriminatory power above
90 per cent and a correctness likelihood
indistinguishable from unity.  Superhumps have
apparently not been reported, but at this $P_{\rm orb}$ 
they are likely to appear in future outbursts.

\subsection{ASAS-SN 15cm} 

This was detected by ASAS-SN on 2015 Jan.~31, at
magnitude 15.77.
The CRTS DR2 light curve has 369 measurements
between 2005 and 2013, and a single outburst to
16.3 mag in 2006; otherwise the source varies
between 18.4 and 18.9.  The SDSS DR12 gives 
$(ugriz) = (19.94, 19.10, 18.58, 18.41, 18.35)$.
SDSS also took a spectrum, which shows late-type
features (in particular MgI $\lambda 5175$ and
NaD absorption) and weak, broad, double-peaked
H$\alpha$ emission.  The SDSS best fit is a spectrum
of type K3.  Our mean spectrum (top panel of 
Fig.~\ref{fig:asn15cm}) appears nearly identical
to the SDSS spectrum, and has nearly the same
absolute flux as well. The H$\alpha$ line is 
double-peaked, with peaks at $\pm 390$ km s$^{-1}$ 
from the line center, and an equivalent width of only 
$\sim 9$ \AA .

We were unable to measure usable emission-line velocities
due to faintness of the source and the weakness of
the H$\alpha$ emission, but the cross-correlation
technique yielded radial velocities for the 
late-type component.  These showed a strong 
modulation near $P_{\rm orb} = 0.209$ d, or 
5.00 hr (Fig.~\ref{fig:asn15cm}, middle panel).  
Differential photometry of the source with the 1.3 m 
telescope and Andor camera quickly revealed a strong 
ellipsoidal modulation (Fig.~\ref{fig:asn15cm}, bottom 
panel).  

The ellipsoidal modulation appears in the CRTS DR2 
data, unambiguously but at low signal-to-noise.
The CRTS data constrain $P_{\rm orb}$ to 0.208466(2) d,
which we adopt.

The spectral decomposition procedure (Section 2) constrains
the spectrum to K2.5 $\pm 2.5$ subclasses, 
in good agreement with the SDSS fit.
This is much warmer than expected at $P_{\rm orb} = 5$ hr;
the tabulation by \citet{kniggedonors} shows that CVs
at this period more typically have secondaries around
type M3.  ASAS-SN 15cm therefore joins a small club
of warm-secondary CVs (see, e.g. \citealt{thorcss1340}, 
\citealt{rebassa14}, and \citealt{thorasn13cl}).

We can constrain the orbital inclination $i$ as follows.
We modeled the ellipsoidal
variation using the light-curve synthesis program described
by \citet{TA05}.  It proved difficult to match the large
amplitude of the observed ellipsoidal modulation, without 
artificially increasing gravity darkening and decreasing 
the disk contribution excessively.  Therefore, $i$ is probably
close to edge-on, but it cannot be too close, because 
the light curve shows no sign of an eclipse of the accretion
disk or white dwarf.  For a mass ratio $q = M_2/M_1 = 0.1$,
which is smaller than likely, the white dwarf (i.e., the
center of the accretion disk) will eclipse
if $i > 79$ degrees \citep{chanan76}. The large-amplitude
ellipsoidal modulation, and the lack of eclipses, constrain 
the plausible range for $i$ to be between about 70 and 
79 degrees.

In this range of inclination, the observed 
$K_2 = 229$ km s$^{-1}$ is consistent with 
white dwarf masses near 0.6 -- 0.7 M$_{\odot}$ and $q \sim 0.6$.
The mass ratio exceeds unity for white dwarf masses 
greater than 1 M$_{\odot}$.  
The secondary-based distance estimation (see Section 2) gives
$2800 \pm 400$ pc.  The Galactic latitude is 30.7 degrees,
so the star is likely to be at least 1 kpc from the Galactic plane.

\subsection{CSS1735+15}

A spectrum of this object was published by \citet{thorskinner}, 
who found H$\alpha$ in emission and noted indications of a 
late-type secondary contribution.  \citet{kato_iv} discovered photometric
modulations in outburst at periods of 0.05436 and 0.05827 d,
which they found puzzling but tentatively interpreted as 
a negative superhump and an orbital period respectively.  

We obtained time series spectroscopy in 2014 June.
The mean spectrum (Fig.~\ref{fig:cvplot7})
shows a strong contribution from a star
with spectral type K4 $\pm 1$ subtype.  Decompositions indicate
that the K star contributes about 80 per cent of the 
$V$-band light; fractions between 55 and 90 per
cent gave acceptable results.

The absorption-line velocities
derived from cross-correlation (and excluding the NaD line, which
can be confused with HeI $\lambda 5876$ emission) shows a
modulation at $P_{\rm orb} = 0.3534(10)$ d, or 8.48 hr, with semi-amplitude
$K = 136(4)$ km s$^{-1}$.  The H$\alpha$ emission line velocities
have lower signal-to-noise but have a modulation consistent with 
this period.  The 8.48-hr period therefore represents the orbital
period of the CV, and is not from some line-of-sight 
interloper.


If we (1) assume the absorption velocity amplitude $K_2$ accurately
reflects the secondary center-of-mass motion (2) assume that the white 
dwarf mass is above 0.6 solar masses, and (3) require the mass
ratio $q = M_2 / M_1 < 1.$, then the orbital inclination 
is less than about 60 degrees.  The phase of the emission velocities
is offset from the absorption by $0.574 \pm 0.017$, while 
the white-dwarf velocity should be offset by a half-cycle;
consequently, we think it unwise to interpret the 
emission line velocities as the white-dwarf motions.  The 
\citet{bk00} models that best match the secondary's spectral
type at this orbital period have $M_2 \sim 0.65$ M$_{\odot}$.

We again estimated a distance using the secondary spectrum
(Section 2).  Because the 
absolute flux measurement through the spectrograph slit is uncertain,
we referred to a single $V$-band acquisition image from the
MDM OSMOS instrument (see
\citet{thorskinner}), which was taken in partly cloudy 
conditions, and calibrated it using $V$ magnitudes
of ten stars in the field taken from APASS 
\citep{henden15}\footnote{APASS, the AAVSO
Photometric All Sky Survey, is described at 
https://www.aavso.org/apass}, yielding $V = 17.9 \pm 0.1$ 
(estimated uncertainty).  For a nominal $M_2 = 0.6 \pm 0.15$ M$_{\odot}$
(used only to estimate the Roche lobe size), the distance came 
to $1830 \pm 330$ pc.  

While the secondary spectral type is normal for this $P_{\rm orb}$,
the superhump-like modulations reported in outburst by \citet{kato_iv}
are unexpected, and their interpretation is not 
clear.  The difference between the frequencies of the two modulations
bears no obvious relationship to the orbital frequency.


\subsection{OT1759+14}

This transient was discovered by ASAS-SN on 2013 April 05, with 
$V = 14.5$, and was designated ASAS-SN 13ab \citep{shappee13}.  
A spectrum taken soon after discovery was consistent with a dwarf nova
outburst.  The transient was also detected by MASTER on 2013 April 11. 
Its location is not covered by the CSS.  The star lies 
15 arcsec from the Einstein X-ray source 2E 1756.8+1451.
The PPMXL catalog \citep{ppmxl}, which includes photographic 
magnitudes from the USNOB catalog \citep{usnob}, gives 
$b1 = 18.57$ and $r1 = 17.16$; a crude transformation
published by John Greaves\footnote{Available at 
http://www.aerith.net/astro/color\_conversion/JG/USNO-B1.0.html} then
gives $V = 17.8 \pm 0.5$.  Our mean spectrum 
(Fig.~\ref{fig:cvplot7}) has a
synthesized $V = 18.1$, but included some data taken
in poor seeing, so we adopt $V = 17.9$ as a rough magnitude
at minimum.

The mean spectrum shows a strong secondary-star contribution
with a spectral type of K5.5 $\pm$ 2 subtypes.  Nascent
molecular bands in the continuum indicate a slightly later
type than the absorption lines, which match best around
K5.  The secondary contributes about half the light 
in the $V$ band, typically giving around $V = 18.6 \pm 0.3$ 
for the secondary contribution.   

The radial velocity of the K star is modulated with a period
of 7.16(2) hr and a semiamplitude $K = 196(14)$ km s$^{-1}$;
the orbit is unambiguous with only 17 velocities taken
over 6 nights.  The best period in the H$\alpha$ emission-line 
velocity matches the absorption period within the mutual
error; we adopted the weighted average, 7.17(2) hr.  

The phase at which the emission lines increase through their
mean velocity is $0.51 \pm 0.04$ in the absorption-line
ephemeris, which is consistent with the phase expected 
if the emission lines trace the white dwarf motion.  
If we take this at face value, the nominal mass ratio
is $M_2 / M_1 = K_1 / K_2 = (72 \pm 15) / (196 \pm 14) = 0.37 \pm 0.08$,
which is smaller than one might expect.  
We have only 17 velocities, and the reliability of
emission line velocities for dynamics is 
questionable, but this does suggest a somewhat more massive
white dwarf than usual, or an undermassive secondary.
Because of this, for the Monte Carlo distance estimate we assumed a 
secondary mass distribution centering on 0.5 M$_{\odot}$, slightly
lower than suggested by the \citet{bk00}
calculations.  The secondary mass enters
only weakly into the calculation, which gave $d = 1600 \pm 400$ pc.
Distances less than 900 pc and greater than 3 kpc are very
unlikely.


\subsection{DDE23}

This object was discovered by \citet{denisenko23} in a 
search for variable stars in the USNO B1.0 catalog.
It lies outside the sky coverage of CSS.  
Nearly all our data, from 2013 June, were taken
using the $1024^2$ `Templeton' detector and hence cover from
just short of HeII $\lambda4686$ to somewhat longward
of H$\alpha$ (see Fig.~\ref{fig:cvplot7}).  
The source was in a relatively 
bright state at that time (synthesized
$V$ = 18.6), showing strong emission lines, 
including HeII $\lambda 4686$ 
over half the strength of H$\beta$, which 
usually indicates a magnetic system.  We
returned to the source in 2013 September, but found
it much fainter and were unable to obtain 
further velocities.  

Although the data are not extensive, the radial 
velocities from 2013 June define a unique 
period of 2.210(2) hr, with a large 
velocity semiamplitude $K = 318(21)$ km s$^{-1}$.
The large $K$ and strong HeII $\lambda 4686$
strongly suggest that this is a polar, or AM Her star.

\subsection{ASAS-SN 13bs}


This object was discovered by ASAS-SN on 2013 August 1.
It is associated with the ROSAT 
X-ray source 1RXS J183008.4+373636
\citep{haakonsen09}.  This part of the sky
is not covered by the CSS, and its outburst
history appears to be unknown.

We observed the source in 2013 September, but could
not determine an unambiguous orbital period.  
In 2014 June we obtained spectra with better
hour-angle coverage and less velocity scatter,
and determined the daily cycle count without
ambiguity. The correct period corresponds to 
the third-best alias in the September data.

The mean spectrum from 2014 June 
(Fig.~\ref{fig:cvplot7}) shows strong, 
incipiently double-peaked emission lines on a blue
continuum; the 2013 September spectrum appears
similar.  H$\alpha$ has an EW of 110 \AA\ and   
a FWHM of 1600 km s$^{-1}$. HeII $\lambda 4686$
is present but much weaker than H$\beta$, and
HeII $\lambda 5411$ is absent despite very good
signal-to-noise, so the system is unlikely
to be magnetic; it is apparently a dwarf 
nova. 

Separate fits to the velocities from the two
observing runs yield consistent periods with
a weighted average of 82.44(11) min.  At this
period, superoutbursts and superhumps can
be expected.


\subsection{DDE14 = 1RXS J185310.0+594509}

This is among the few CVs in this paper that were not discovered
by a variability survey, though soon after it was first recognized
it did turn up in the ASAS-SN survey.  It was identified by
\citet{denisenko11}, who found it by searching for blue and variable
stars in the USNO-B catalog \citep{usnob} in the vicinity of unidentified
ROSAT sources.  They classified the star as a CV
and identified it with 1RXS J185310.0+594509.
ASAS-SN detected an outburst
on 2013 August 13, at $V = 15.18$.  The relatively
sparse CRTS light curve (67 measurements in 8 
years) did not catch any outburst, mostly showing the source 
between 18th and 19th mag, with one point at 19.87. 
\citet{katovii} report a superhump period of 0.059521(32) d, 
or 85.71(5) min, confirming that this is an SU UMa star.




The spectrum (Fig.~\ref{fig:cvplot8})
shows the strong, double-peaked emission
lines characteristic of a high-inclina\-tion dwarf nova
at minimum light.  H$\alpha$ has an EW of 90 \AA\ and a
FWHM of 2000 km s$^{-1}$; its two peaks are separated
by 1050 km s$^{-1}$.  
The radial velocities unambiguously indicate 
$P_{\rm orb} = 83.89(14)$ min.  This is well within the 
range expected from the observed superhump period
\citep{unveils}.


\subsection{CSS2133+19}


This is one of 18 dwarf novae found and classified by 
\citet{drakevars14}, as byproducts of a search
for periodic variables in the CSS data.  The CSS light
curve shows a minimum varying from 17.0 to 17.5 mag, and at least
five distinct outbursts between 2006 and 2013, reaching
a maximum near 14.4 mag.  \citet{drakevars14} flagged
this and three of the other CVs that were in the SDSS
footprint as having a color excess suggesting a strong
contribution from the companion star.  

The mean spectrum (Fig.~\ref{fig:cvplot8})
shows a significant contribution from 
a K5 ($\pm$ about 1.5 subclass) secondary, which 
contributes about 40 \% of the 
light in the V band.  The magnitude is uncertain;
the synthesized magnitude is $V = 17.3$, but the 
SDSS DR12 magnitudes and colors imply $V \sim 17.6$, and
the mean CSS magnitude around minimum is $V \sim 17.2$. 
For purposes of estimating the distance, we take the 
magnitude at minimum light to be $V = 17.3 \pm 0.5$.

The radial velocities of the absorption and 
emission yield nearly identical periods, with a
weighted average of 9.025(12) hr.  The secondary's
spectral type is typical for CVs near this period
\citep{kniggedonors}.  
The Monte Carlo procedure distance estimation procedure
described in Section 2 yields a distance of 
1890 (+610, $-$440) pc.  Distances less than 1 kpc and
greater than $\sim 3$ kpc are unlikely.  At these
Galactic coordinates ($l = 71$ and $b = -23.4$ degrees)
and a distance of 1890 pc, the source would lie 750 pc 
from the Galactic plane.



\subsection{CSS2156+19} 

We have no spectra of this object, but 
\citet{szkody14} obtained a blue spectrum
which shows double-peaked Balmer emission indicating 
a high inclination.
The CRTS listing notes an eclipse, so we
obtained differential time-series photometry with 
the 1.3 m telescope and 
Andor camera on three nights in 2014 October.  The 
light curves (Fig.~\ref{fig:css2156lightcurves}) show
an eclipse with a depth of $\sim$ 2 mag on a period
near 102 minutes, along with a strong pre-eclipse
hump.  At minimum light the object was near our 
limiting magnitude, so the eclipse depth is not
measured accurately, but it appears to be at least 
2 mag.  Our 2014 October data constrain the period to be 
0.070922(9) d.  To refine this, we folded the 
CRTS archival data on periods near this,
and examined the resulting folded light curve while
gradually shifting the fold period.  An apparent
eclipse emerged from the data at $P = 0.0709291(1)$
(see Fig.~\ref{fig:crtsphot}).
A single eclipse kindly obtained with the 1.3m
telescope in 2015 December by K. Bakowska confirms
the long-term ephemeris, which is
$$\hbox{BJD mid} = 2456941.7307(4) + 0.0709291(1) E,$$
where $E$ is the cycle count.

\subsection{CSS2227+28}

The Catalina survey discovered this object on 2009 May 31.  The
CRTS light curve shows only two outbursts from a typical 
quiescent magnitude near 18.  \citet{thorskinner} published
a spectrum covering the H$\alpha$ emission line, which was
single-peaked with an EW of 115 \AA .  The present mean
spectrum (Fig.~\ref{fig:cvplot8})
shows H$\alpha$ even stronger, with 160 \AA\ equivalent
width and a FWHM close to 1000 km s$^{-1}$.  
The synthesized $V$ magnitude is $\sim18.6$, 
consistent with the quiescent state.  The spectrum
appears typical for a quiescent dwarf nova.

The radial velocities indicate a period near 131.0(3) min, 
with little ambiguity in the daily cycle count.  This 
is just at the lower edge of the so-called 2-to-3 hour 
gap in the CV period distribution.

\subsection{CSS2319+33}


This object is apparently the optical counterpart of the 
ROSAT X-ray source 1RXS J231909.9+ 331544.  The CRTS
light curve shows irregular variation from 16.3 to
19.5 mag between 2005 and 2014, with the bulk
of the measurements between 17th and 18th magnitude.
Discrete outbursts are not seen, suggesting that this
is a novalike variable rather than a dwarf nova.

The mean spectrum (Fig.~\ref{fig:cvplot8}) 
shows strong, single-peaked emission
lines.  H$\alpha$ has an EW of 110 \AA\ and a 
FWHM of 1200 km s$^{-1}$.  HeII $\lambda 4686$ (near the edge of the 
spectral coverage) appears weakly, and HeII $\lambda 5411$
is not apparent, so this is unlikely to be a magnetic
system.  The H$\alpha$ radial velocity shows a clear
modulation at 3.508(5) hr; cataclysmics in this period
range tend to be novalike variables rather than 
dwarf novae. Many CVs in this period range show
complex phase-dependent changes in the line profile,
associated with the SW Sex phenomenon \citep{thorstensen91} 
but this was not apparent in a phase-averaged
greyscale representation of the spectra (not shown).


\subsection{CSS2335+12}

This source was found by the Catalina survey 
on 2013 December 14.  Its CSS light curve shows 
irregular variability around a mean
of roughly 19th magnitude, with excursions as bright 
as 17.3 mag and fainter than 19.5 mag. 
A single measurement in 2013 June puts it at 20.3 mag.  
The CSS listing notes it as a possible polar, or AM Her
star.

The mean spectrum (Fig.~\ref{fig:cvplot9}) 
shows HeII $\lambda 4686$ with a strength
comparable to H$\beta$, as well as HeII $\lambda 5411$.
The lines are narrow with broad bases, which together 
with the high excitation support the polar classification.
There is a contribution from a late-type star, which is 
almost certainly the secondary star.  The late-type
component cannot be classified to high precision --
acceptable decompositions are possible from 
M1 to M4.  The secondary's contribution gave 
a distance estimate (Section 2) of $1400 \pm 500$ pc.

We have radial velocities from two adjacent nights
that define an unambiguous period of 3.884(17) hr.
Fig.~\ref{fig:trails} shows a phase-averaged
greyscale representation of the spectrum.   The
H$\alpha$ line shows a sharp component with a clear 
velocity modulation, and a diffuse component 
with a greater amplitude that is offset in phase,
a pattern commonly seen in polars.  The sharp
component is probably from the heated face of the
secondary star, and the diffuse component from the 
accretion stream near the white dwarf.  
In the greyscale representation the 
HeII $\lambda 5411$ line (not shown) shows only the diffuse, 
large-amplitude component, consistent with an
origin in the accretion column.  Although $\lambda 5411$ is 
weaker than HeII $\lambda 4686$, it is more
distinct in the greyscale because of the better
instrumental sensitivity at the longer wavelength.
 
\section{Summary and Discussion}

In the last column of Table~\ref{tab:star_info}, 
we give the subtype of each of the objects, and 
briefly note other notable characteristics.  The majority
of the stars were found in variability surveys, so
it is not surprising that most are dwarf novae, or
U Geminorum stars; a few are listed as `UG:' because
their outburst history is not known.  Twenty of the 
known or suspected dwarf novae have $P_{\rm orb}$
shorter than $\sim 3$ hr, the period range in which 
superoutbursts and superhumps are found; however,
only four of the 20 have known superhump periods.  

A superhump period would be especially desirable for
SDSS1029+48, which appears to be a low
mass-transfer rate system, with an apparent
white dwarf component in its spectrum, and broad,
double-peaked lines that show little orbital motion.
Its orbital period of 91 min,
(a cycle-count alias is unlikely) is 
significantly longward of the $\sim 75$-min period minimum
for CVs with `normal' hydrogen-rich secondaries,
so it is a candidate `post-bounce' system.   
If this is the case, it should have 
a superhump period excess that is anomalously
small for its orbital period.

While most of the sample are either dwarf novae or
resemble them spectroscopically, a few are not.
Three objects appear to be polars, or AM Her stars  --
CSS0357+10 (already classified by \citealt{schwope12}),
DDE23, and CSS2335+12.  Two others, SDSS1429+00
and CSS2319+33, appear to be novalike variables 
in the period range frequented by SW Sex stars.

We find eclipses in four of the objects.  
The most unusual
of these is ASAS-SN 15aa, which has a relatively long
$P_{\rm orb}$ of 9.01 hr, a large secondary-star contribution,
a sizeable ellipsoidal variation, and a distinctive,
asymmetric, shallow eclipse.  Another, ASAS-SN 15cw,
eclipses deeply on a 114-min period and shows a spectacular
rotational disturbance in its emission line velocities.

In six of the stars we detect late-type secondaries
well enough to classify them and estimate distances.
In most of them the secondary's spectral type is
as expected given $P_{\rm orb}$, but the secondary
in ASAS-SN 15cm is anomalously warm for its 5.0-hr
period. That system also shows a strong photometric
modulation that appears to be mostly ellipsoidal.

Two objects, ASAS-SN 14dx and ASAS-SN 14ag, 
show significant proper motions and appear to be
relatively nearby CVs that have escaped attention 
until their recent discovery.

\citet{unveils} showed that the sample of CVs
discovered by SDSS finally revealed the long-sought
`spike' in the period distribution just above the 
minimum period for hydrogen-burning CVs.  It 
is interesting to ask whether, in an analogous fashion,
the present sample might show any systematic difference 
from the sample of previously-known CVs. 

For the period distribution, the answer to this
is evidently ``no''.  Fig.~\ref{fig:cdf} shows
a comparison of the cumulative distribution functions
(CDFs) of the periods
presented in this paper, most of which appear here
for the first time, and the periods of 
999 CVs and related objects with well-determined
periods listed in version 7.23 of the 
Ritter-Kolb catalog (\citealt{rkcat}; hereafter
RKcat).  The distributions are remarkably similar, 
except for
the small tail of very short period systems (mainly
AM CVn systems) present in
RKcat.  Both CDFs show a predominance of 
systems below the 2-3 hr `gap', and show 
a distinct change of slope in the gap.
The present sample does 
have three systems with $P_{\rm orb} > 8$ hr,
whereas fewer might be expected given the 
RKcat distribution, but the differences between
the two distributions are not statistically significant
by the Kolmogorov-Smirnov test.  

It is interesting that so many new CVs are still being
turned up.  \citet{breedt14} argue that the 
Catalina survey has already discovered most of the
high-accretion rate dwarf novae within its survey
area.  In our sample, 30 of the 35 objects were 
discovered by the ASAS-SN, Catalina, or MASTER surveys,
and therefore have well-defined discovery dates. 
Of these, 5 were discovered in 2015, 8 in 2014, 
6 in 2013, and 11 before that.  These numbers
are small, but do not show an obvious
decline in the discovery rate.  It may take some
time before we can be confident that even the 
dwarf nova sample is close to complete.

\section{Acknowledgements}

We gratefully acknowledge support from NSF grant
AST-1008217, and thank the MDM staff for observing 
assistance.  We also thank Karolina Bakowska of the 
Nicolas Copernicus Astronomical Center for obtaining
an observation of CSS 2156+19.

This paper uses data from the
Catalina Sky Survey and Catalina Real Time Survey; the CSS is 
funded by the National Aeronautics and Space
Administration under Grant No. NNG05GF22G issued through the Science
Mission Directorate Near-Earth Objects Observations Program.  The CRTS
survey is supported by the U.S.~National Science Foundation under
grants AST-0909182.  This research has made use of the APASS database, 
located at the AAVSO web site. Funding for APASS has been provided by 
the Robert Martin Ayers Sciences Fund.
This paper uses data from the Sloan Digital Sky Survey.
Funding for SDSS-III has been provided by the Alfred P. 
Sloan Foundation, the Participating Institutions, the 
National Science Foundation, and the U.S. Department of 
Energy Office of Science. The SDSS-III web site is http://www.sdss3.org/.
SDSS-III is managed by the Astrophysical Research Consortium for the 
Participating Institutions of the SDSS-III Collaboration.

Facilities: \facility{Hiltner, McGraw-Hill}

\clearpage

\begin{deluxetable}{lrrrl}
\tabletypesize{\footnotesize}
\tablewidth{0pt}
\tablecolumns{5}
\tablecaption{List of Objects}
\tablehead{
\colhead{Short Name} &
\colhead{Other Name} & 
\colhead{$\alpha_{\rm 2000}$} &
\colhead{$\delta_{\rm 2000}$} & 
\colhead{Type} \\
\colhead{} &
\colhead{} &
\colhead{[h m s]} &
\colhead{[$^{\circ}$ $'$ $''$]} &
\colhead{} 
}
\startdata
CSS0015+26 & CSS090918:001538+263657 &  0 15 38.26 & +26 36 56.8 & UG(SU)\\
ASAS-SN 14dr & & 0 36 40.27 & +23 08 33.2 & UG(SU)\\
CSS0133+38 & CSS091027:013309+383217 &  1 33 08.72 & +38 32 17.2  &UG(SU)\\
CSS0143+26 & CSS141029:014305+263833 &  1 43 04.69 & +26 38 33.0  &UG(SU)\\
OT0150+37  & MASTER J015017.0+375614 &  1 50 16.19 & +37 56 19.0  &UG: \\ 
CSS0208+37 & CSS101207:020804+373217 &  2 08 04.23 & +37 32 16.6  &UG(SU)\\
ASAS-SN 14dx & & 2 34 27.59 &  $-$04 54 34.3  &UGSU; PM\\
CSS0357+10 & CSS091109:035759+102943 &  3 57 58.67 & +10 29 42.9& Polar  \\
CSS0500+13 & CSS151201:050027+133420 &  5 00 27.24 & +13 34 20.0& UG; 2ndary \\ 
OT0647+49  & MASTER J064725.70+491543.9 &  6 47 25.70 & +49 15 43.9& UGSU \\
ASAS-SN 14kj & & 8 02 34.39 & $-$00 09 40.2 & UG: (SU) \\   
ASAS-SN 15cw & & 8 08 18.98 & +00 59 00.1 & UG:; ecl\\  
ASAS-SN 14ag & & 8 13 18.52 & $-$01 03 29.5 &UGSU; ecl; PM\\
ASAS-SN 14ak & & 9 50 57.04 & $-$29 46 46.0 &UG(SU)\\ 
CSS1028$-$16 & CSS140309:102844-161303 & 10 28 43.87 & $-$16 13 03.4& UG:(ZC?)\\ 
SDSS1029+48 & SDSS J102905.21+485515.23 & 10 29 05.24 & +48 55 15.2& UG:(WZ?) \\ 
ASAS-SN 15aa  & & 10 49 25.91 & $-$21 47 35.9& UG; ecl; 2ndary  \\
CSS1055+09 & CSS080130:105550+095621 &  10 55 50.08 & +09 56 20.5&UG; 2ndary \\ 
CSS1211$-$08 & CSS110205:121119-083957 & 12 11 19.13 & $-$08 39 57.1& UG(SU) \\ 
SDSS1429+00 & SDSS J142958.15+000319.6 & 14 29 58.15 & +00 03 19.6& NL \\
ASAS-SN 15dl & & 16 22 00.39 & $-$22 41 17.0 & UG\\
CSS1631+10 & CSS080505:163121+103134 & 16 31 20.89 & +10 31 33.9 & UG \\
CSS1702+55 & CSS130416:170209+551827 & 17 02 08.59 & +55 18 26.7 & UG(SU) \\
OT1727+38 & MASTER J172758.09+380021.5 & 17 27 58.14 & +38 00 22.5 & UG(SU)\\ 
ASAS-SN 15cm & & 17 31 02.21 & +34 26 33.1 & UG; 2ndary\\ 
CSS1735+15 & CSS110623:173517+154708 & 17 35 16.91 & +15 47 08.4 & UG; 2ndary \\
OT1759+14  & MASTER J175908.60+145130.2 & 17 59 08.60 & +14 51 30.2 & UG; 2ndary \\
DDE23      & & 18 19 33.12 & +58 06 26.3  & Polar\\
ASAS-SN 13bs  & & 18 30 08.84 & +37 36 41.6 & UG: (SU) \\
DDE14      & 1RXS J185310.0+594509 & 18 53 09.61 & +59 45 06.8 & UGSU \\
CSS2133+19 & CSSJ213319.4+190155 & 21 33 19.47 & +19 01 55.0 & UG; 2ndary \\
CSS2156+19 & CSS090622:215636+193242 & 21 56 36.32 & +19 32 41.9 & UG(SU); ecl \\
CSS2227+28 & CSS090531:222724+284404 & 22 27 24.49 & +28 44 03.8 & UG(SU)\\
CSS2319+33 & CSS111021:231909+331540 & 23 19 09.18 & +33 15 39.8 & NL \\
CSS2335+12 & CSS131214:233546+123448 & 23 35 45.95 & +12 34 48.4 & Polar \\
\enddata
\tablecomments{Coordinates are for the most part from SDSS or the 
PPMXL catalog \citep{ppmxl}.  In the last column, the notation is as 
follows: UG -- U Gem star, or dwarf nova, with a colon indicating that this
is uncertain because outbursts have not been reported; UG(SU) -- 
U Gem star in the period range in which superhumps
may be expected, but for which they have not been reported; UGSU -- SU UMa
star with measured superhump period; UG:(WZ?) -- apparently 
slow mass transfer dwarf nova without reported outbursts; UG:(ZC?) -- possible
Z Cam-type dwarf nova; Polar -- AM Her
star; PM -- significant proper motion; 2ndary -- secondary star detected in 
spectrum; ecl -- eclipses are detected.}
\label{tab:star_info}
\end{deluxetable}

\begin{deluxetable}{lrrrrr}
\tablecolumns{6}
\tablewidth{0pt}
\tablecaption{Radial Velocities}
\tablehead{
\colhead{Star} &
\colhead{Time\tablenotemark{a}} &
\colhead{$v_{\rm emn}$} &
\colhead{$\sigma$} & 
\colhead{$v_{\rm abs}$} &
\colhead{$\sigma$} \\
\colhead{SDSS J} &
\colhead{} &
\colhead{[km s$^{-1}$]} &
\colhead{[km s$^{-1}$]} &
\colhead{[km s$^{-1}$]} &
\colhead{[km s$^{-1}$]} \\
}
\startdata
CSS0015+26  &  2456942.9461  &     12  &     8  &  \nodata  &  \nodata \\
CSS0015+26  &  2456942.9535  &     11  &     9  &  \nodata  &  \nodata \\
CSS0015+26  &  2456942.9608  &     52  &    10  &  \nodata  &  \nodata \\ 
CSS0015+26  &  2456943.6660  &     47  &     8  &  \nodata  &  \nodata \\
CSS0015+26  &  2456943.6733  &     28  &     9  &  \nodata  &  \nodata \\ 
CSS0015+26  &  2456943.6807  &     14  &     9  &  \nodata  &  \nodata \\
CSS0015+26  &  2456943.6880  &     -2  &    10  &  \nodata  &  \nodata \\
CSS0015+26  &  2456943.6953  &     -8  &    10  &  \nodata  &  \nodata \\
\enddata \\
\tablenotetext{a}{Barycentric Julian Date of mid-exposure.  The time base is UTC.}
\tablecomments{Table \ref{tab:velocities} is published in its entirety in the electronic 
edition of The Astronomical Journal, A portion is shown here for guidance regarding its form and content.}
\label{tab:velocities}
\end{deluxetable}

\clearpage

\begin{deluxetable}{lllrrcc}
\tablecolumns{7}
\footnotesize
\tablewidth{0pt}
\tablecaption{Fits to Radial Velocities}
\tablehead{
\colhead{Data set} & 
\colhead{$T_0$\tablenotemark{a}} & 
\colhead{$P$} &
\colhead{$K$} & 
\colhead{$\gamma$} & 
\colhead{$N$} &
\colhead{$\sigma$\tablenotemark{b}}  \\ 
\colhead{} & 
\colhead{} &
\colhead{(d)} & 
\colhead{(km s$^{-1}$)} &
\colhead{(km s$^{-1}$)} & 
\colhead{} &
\colhead{(km s$^{-1}$)} \\
}
\startdata
CSS0015+26 & 56944.8652(11) & 0.10150(6) &  77(4) & $-36(3)$ & 28 &  11 \\[1.2ex]
ASAS-SN 14dr & 57317.9693(15) & 0.06655(9) &  59(8) & $-67(6)$ & 49 &  27 \\[1.2ex]
CSS0133+38 & 56944.968(3) & 0.07497(13) &  47(10) & $-4(8)$ & 41 &  26 \\[1.2ex]
CSS0143+26 & 57002.655(2) & 0.0646(2) &  47(9) & $-33(6)$ & 36 &  21 \\[1.2ex]
OT0150+37  & 56597.871(3) & 0.1481(3) &  84(10) & $-31(7)$ & 58 &  32 \\[1.2ex]
CSS0208+37 & 56300.6700(16) & 0.05845(9) &  63(11) & $-55(8)$ & 58 &  36 \\[1.2ex]
ASAS-SN 14dx & 56945.8452(7) & 0.05756(6) &  41(3) & $ 8(2)$ & 43 &  10 \\[1.2ex]
CSS0357+10 & 56298.6443(7) & 0.07918(6) &  320(19) & $ 11(13)$ & 33 &  62 \\[1.2ex] 
CSS0500+13 & 57440.7942(9) & 0.15465(3) &  126(5) & $-6(3)$ & 46 &  16 \\[1.2ex]
OT0647+49  & 57000.9276(14) & 0.06554(12)\tablenotemark{c} &  78(10) & $-30(7)$ & 55 &  33 \\[1.2ex]
ASAS-SN 14kj & 57433.935(3) & 0.0906(2) &  82(18) & $ 14(13)$ & 45 &  42 \\[1.2ex]
ASAS-SN 15cw & 57430.9527(18) & 0.0791258(4)\tablenotemark{d} &  78(12) & $-40(8)$ & 13 &  22 \\[1.2ex]
ASAS-SN 14ag & 57402.8461(17) & 0.060310651(2)\tablenotemark{d} &  100(19) & $ 31(13)$ & 44 &  43 \\[1.2ex]
ASAS-SN 14ak & 57433.7837(14) & 0.08869(7) &  73(7) & $ 4(5)$ & 33 &  17 \\[1.2ex]
CSS1028$-$16 & 57460.869(4) & 0.1897(6) &  102(12) & $-3(9)$ & 53 &  40 \\[1.2ex]
SDSS1029+48 & 57430.7432(18) & 0.063421(8) &  22(4) & $-10(3)$ & 58 &  13 \\[1.2ex]
ASAS-SN 15aa [abs.] & 57109.974(3) & 0.375540(2)\tablenotemark{e} &  175(7) & $-23(5)$ & 38 &  23 \\
ASAS-SN 15aa [emn.] & 57109.788(7) & \nodata &  154(14) & $-66(11)$ & 34 &  55 \\[1.2ex]
CSS1055+09 & 57436.0463(16) & 0.1624(5) &  116(6) & $-75(5)$ & 36 &  16 \\[1.2ex]
CSS1211$-$08 & 57438.840(2) & 0.07913(19) &  45(8) & $-26(6)$ & 47 &  22 \\[1.2ex]
SDSS1429+00 & 57460.7595(16) & 0.14833(17) &  172(14) & $-28(9)$ & 53 &  42 \\[1.2ex]
ASAS-SN 15dl & 57142.780(5) & 0.2288(9) &  82(12) & $-5(8)$ & 49 &  28 \\
(alternate)  & 57142.741(8) & 0.2976(18) &  80(14) & $-6(10)$ & 49 &  31 \\[1.2ex]
CSS1631+10 & 55323.9928(19) & 0.06265(11)\tablenotemark{c} &  60(12) & $-35(8)$ & 115 &  38 \\[1.2ex]
CSS1702+55 & 56453.6424(16) & 0.06931(15) &  35(5) & $-27(4)$ & 33 &  12 \\[1.2ex]
OT1727+38  & 57554.7680(15) & 0.05704(4) &  53(8) & $-25(6)$ & 56 &  27 \\[1.2ex]
ASAS-SN 15cm & 57553.8461(11) & 0.208466(2)\tablenotemark{f} & 229(7) & $ 21(5)$ & 21 &  18 \\[1.2ex]
CSS1735+15 [abs.] & 56827.9769(15) & 0.3534(10)\tablenotemark{g} &  136(4) & $-35(3)$ & 27 &  10 \\
CSS1735+15 [emn.]  & 56827.826(6) & \nodata   &  101(15) & $-66(9)$ & 27 &  29 \\[1.2ex]
OT1759+14 [abs.] & 56830.778(4) & 0.2987(8)\tablenotemark{h} &  196(14) & $ 45(11)$ & 17 &  23 \\ 
OT1759+14 [emn.] & 56830.930(12) & \nodata &  72(15) & $ 39(12)$ & 17 &  24 \\[1.2ex]
DDE23      & 56453.8746(9) & 0.09207(9) & 318(21) & $ 36(14)$ & 19 &  41 \\[1.2ex] 
ASAS-SN 13bs & 56551.7588(16) & 0.05725(8)\tablenotemark{c} &  45(8) & $-68(6)$ & 53 &  18 \\[1.2ex]
DDE14      & 56831.9706(16) & 0.05826(9) &  73(15) & $-22(10)$ & 33 &  33 \\[1.2ex]
CSS2133+19 [abs.] & 56942.727(4) & 0.3761(5)\tablenotemark{h} &  151(8) & $ 52(7)$ & 18 &  19 \\
CSS2133+19 [emn.] & 56942.925(4) & \nodata  &  120(7) & $10(5)$ & 20 &  15 \\[1.2ex]
CSS2227+28 & 56944.6401(16) & 0.09101(18) &  48(5) & $-27(4)$ & 27 &  12 \\[1.2ex]
CSS2319+33 & 56551.6415(17) & 0.14616(19) &  96(8) & $10(5)$ & 31 &  18 \\[1.2ex]
CSS2335+12 & 57001.7433(19) & 0.1618(7) &  169(13) & $-25(9)$ & 37 &  31 \\
\enddata
\tablecomments{Parameters of least-squares sinusoid fits to the radial
velocities, of the form $v(t) = \gamma + K \sin(2 \pi(t - T_0)/P$.}
\tablenotetext{a}{Heliocentric Julian Date minus 2400000.  The epoch is chosen
to be near the center of the time interval covered by the data, and
within one cycle of an actual observation.}
\tablenotetext{b}{RMS residual of the fit.}
\tablenotetext{c}{Period given is the weighted mean of periods from separate
fits to the velocities from the individual observing runs.}
\tablenotetext{d}{Period fixed at value derived from eclipses.}
\tablenotetext{e}{For both the absorption and emission fits, the 
period was held fixed at the value derived from the ellipsoidal
variations in the archival CRTS light curve.}
\tablenotetext{f}{Period fixed at the value derived from the
archival CRTS light curve.}
\tablenotetext{g}{Period for emission held fixed at the value derived from
the absorption lines.}
\tablenotetext{h}{Period was fixed at the weighted mean of the absorption- and emission-line
periods.}
\label{tab:parameters}
\end{deluxetable}

\clearpage

\begin{deluxetable}{rrrr}
\tabletypesize{\footnotesize}
\tablewidth{0pt}
\tablecolumns{4}
\tablecaption{Eclipse Timings}
\tablehead{
\colhead{$E$} & 
\colhead{$T$} &
\colhead{$O-C$} &
\colhead{Instrument} \\ 
}
\startdata
\multicolumn{4}{l}{ASAS-SN 15cw :}\\
\multicolumn{4}{l}{$2457431.7674(1) + 0.0791258(4) E$} \\[1.0ex]
$-1$ & 7431.6881 & $-17$ & 1.3m+Andor\\
$0$ & 7431.7675 & $+8$ &  1.3m+Andor\\
$11$ & 7432.6377 & $-12$ &  1.3m+Andor\\
$12$ & 7432.7171 & $10$ &  1.3m+Andor\\
$27$ & 7433.9039 & $7$ &  1.3m+Andor\\
$112$ & 7440.6295 & $7$ & 1.3m+Andor \\
$391$ & 7462.7056 & $-2$ & 1.3m+slit \\[2ex]
\multicolumn{4}{l}{ASAS-SN 14ag:}   \\
\multicolumn{4}{l}{$2457416.7385(1) + 0.060310651(2)E$} \\[1.0ex]
$0$ & 7416.7386 & $9$ & 1.3m+Andor\\
$1$ & 7416.7987 & $-10$ & 1.3m+Andor\\
$2$ & 7416.8591 & $-2$ & 1.3m+Andor\\
$34$ & 7418.7892 & $12$ & 1.3m+Andor\\
$35$ & 7418.8495 & $11$ & 1.3m+Andor\\[2ex]
\multicolumn{4}{l}{ASAS-SN 15aa:}  \\
\multicolumn{4}{l}{$2457109.977(3) + 0.375540(2)E$} \\[1.0ex]
$-6$ & 7107.7234 & $-31$ & 1.3m+slit\\
$2$ & 7110.7286 & 45 & 1.3m+slit\\
$34$ & 7122.7460 & 55 & 1.3m+Andor\\
$783$ & 7404.0242 & $-54$ & 1.3m+Andor \\
$788$ & 7405.9014 & $-97$ & 1.3m+Andor\\[2ex]
\multicolumn{4}{l}{CSS2156+19:} \\
\multicolumn{4}{l}{$2456941.7307(4) + 0.0709291(1)E$} \\[1.0ex]
0    & 6941.73075 & 4 & 1.3m+Andor\\
27   & 6943.64588 & 8& 1.3m+Andor\\
28   & 6943.717   & 25 & 1.3m+Andor\\
29   & 6943.7878  & 13 & 1.3m+Andor\\
41   & 6944.63839 & $-35$& 1.3m+Andor\\
42   & 6944.70959 & $-11$& 1.3m+Andor\\
6075\tablenotemark{a} & 7372.6244  & 0 & 1.3m+Andor\\ 
\enddata
\tablecomments{Column 1: Integer cycle count $E$
in the quoted ephemeris. Column 2: the observed time of mid-eclipse, 
given as the barycentric Julian Date minus 2,450,000.  Column 3: 
Observed timing, minus the time predicted by the quoted ephemeris,
in seconds. Column 4: Telescope and instrument; {\it slit} means
that the Andor camera was viewing the spectrograph slit jaws.
}
\tablenotetext{a}{This observation was kindly obtained by 
Karolina Bakowska.}
\label{tab:eclipsetimings}
\end{deluxetable}

\clearpage

\addtocounter{figure}{-1}
\begin{figure}
\caption{Average spectra, periodograms, and folded velocity
curves for CSS0015+26, ASAS-SN 14dr, CSS0133+38, and CSS0143+26.
The vertical scales, unlabeled to save space, are (1)
for the spectra, $f_\lambda$ in units of $10^{-16}$ erg
s$^{-1}$ cm$^{-2}$ \AA$^{-1}$; (2) for the periodograms,
$1 / \chi^2$ (dimensionless); and (3) for the radial velocity
curves, barycentric radial velocity in km s$^{-1}$.  
When two traces are shown in the spectral plot, the lower
trace shows the average spectrum minus a late-type 
spectrum scaled to match the spectrum of the secondary star.
In cases where velocities are from more than one observing run, 
the periodogram is labeled with the word ``peaks'', because the 
curve shown formed by joining local maxima in the 
full periodogram with straight lines. This suppresses fine-scale
ringing due to the unknown number of cycle counts between runs. 
The folded velocity curves all show the same data plotted over
two cycles for continuity, and the best-fit sinusoid (see 
Table \ref{tab:parameters}) is also plotted.  The velocities shown
are normally H$\alpha$ emission velocities.  Secondary-star
cross-correlation velocities, when available, are shown in 
red, together with the best-fit sinusoid.  The error bars
for the emission lines are computed by propagating the estimated 
noise in the spectrum through the measurment, and hence do not include
jitter due to line profile variations.
}
\end{figure}
 
\begin{figure}
\includegraphics[height=21.5 cm,trim = 2.2cm 2.0cm 2cm 2.8cm,clip=true]{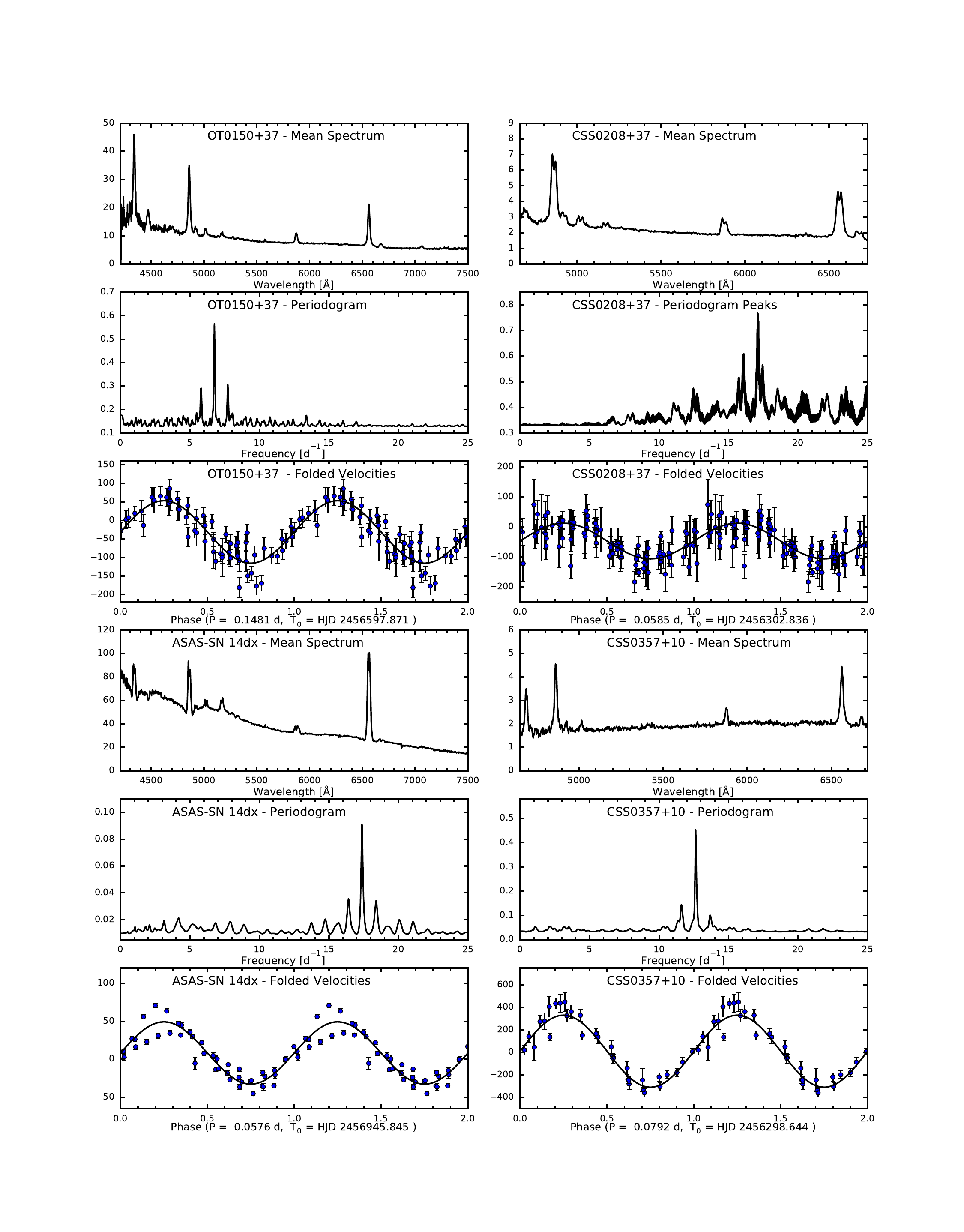}
\caption{Similar to Fig.~\ref{fig:cvplot1}, but for 
OT O150+37, CSS 0208+37, ASAS-SN 14dx, and CSS 0357+10.
}
\label{fig:cvplot2}
\end{figure}

\begin{figure}
\includegraphics[height=21.5 cm,trim = 2.2cm 2.0cm 2cm 2.8cm,clip=true]{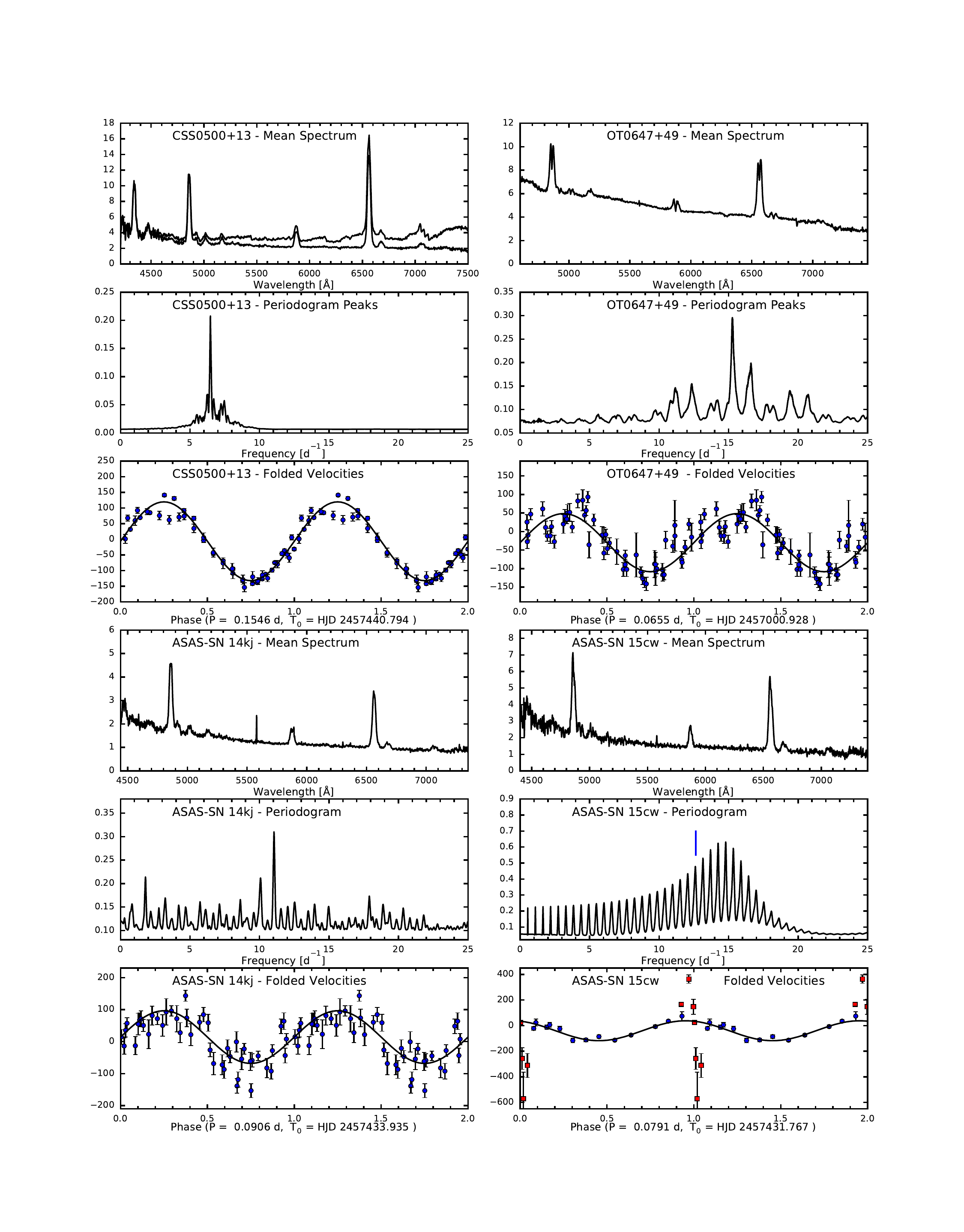}
\caption{(Caption on next page.)}
\label{fig:cvplot3}
\end{figure}
\addtocounter{figure}{-1}
\begin{figure}
\caption{Similar to Fig.~\ref{fig:cvplot1}, but for 
CSS0500+13, OT 0647+49, ASAS-SN 14kj, and ASAS-SN 15cw.  For CSS0500+13, the 
upper trace shows the mean spectrum, and the lower trace shows the result 
of subtracting a scaled spectrum of Gliese 388, an M3.5 dwarf.  For the 
folded velocities on ASAS-SN 15cw, the zero of phase is chosen to correspond
to the eclipse.  The red points show the pronounced rotational disturbance
near eclipse and were not included in the sinusoid fit.  The period for this
star was inferred from eclipses, rather than the velocity periodogram; the 
vertical line in the periodogram plot indicates the period.
}
\end{figure}

\begin{figure}
\includegraphics[height=21.5 cm,trim = 2.2cm 2.0cm 2cm 2.8cm,clip=true]{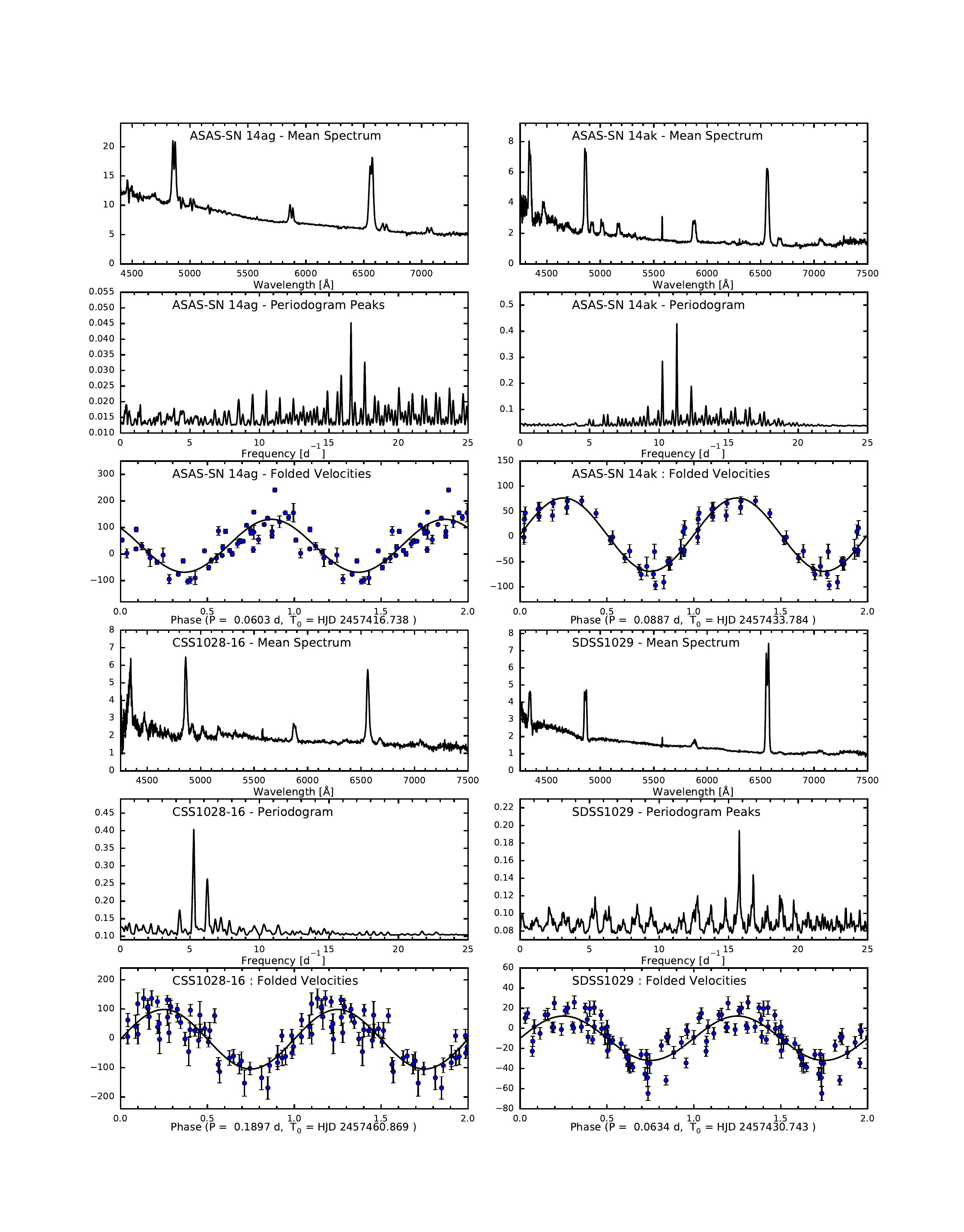}
\caption{Similar to Fig.~\ref{fig:cvplot1}, but for 
ASAS-SN 14ag, ASAS-SN 14ak, CSS 1028$-$16, and SDSS 1029.}
\label{fig:cvplot4}
\end{figure}

\begin{figure}
\includegraphics[height=21.5 cm,trim = 2.2cm 2.0cm 2cm 2.8cm,clip=true]{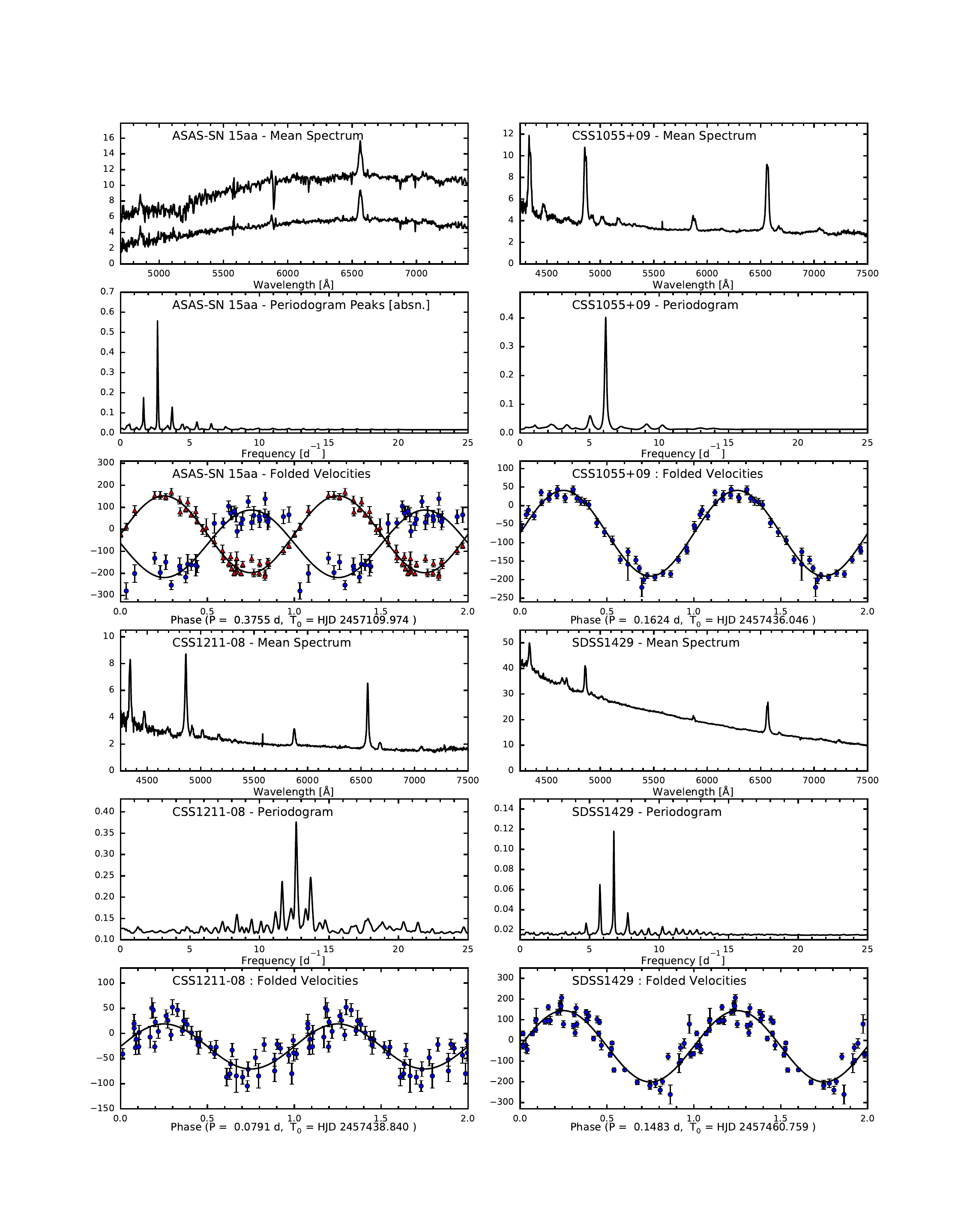}
\caption{Similar to Fig.~\ref{fig:cvplot1}, but for 
ASAS-SN 15aa, CSS1055+09, CSS1211$-$08, and SDSS1429.
}
\label{fig:cvplot5}
\end{figure}
 
\begin{figure}
\includegraphics[height=21.5 cm,trim = 2.2cm 2.0cm 2cm 2.8cm,clip=true]{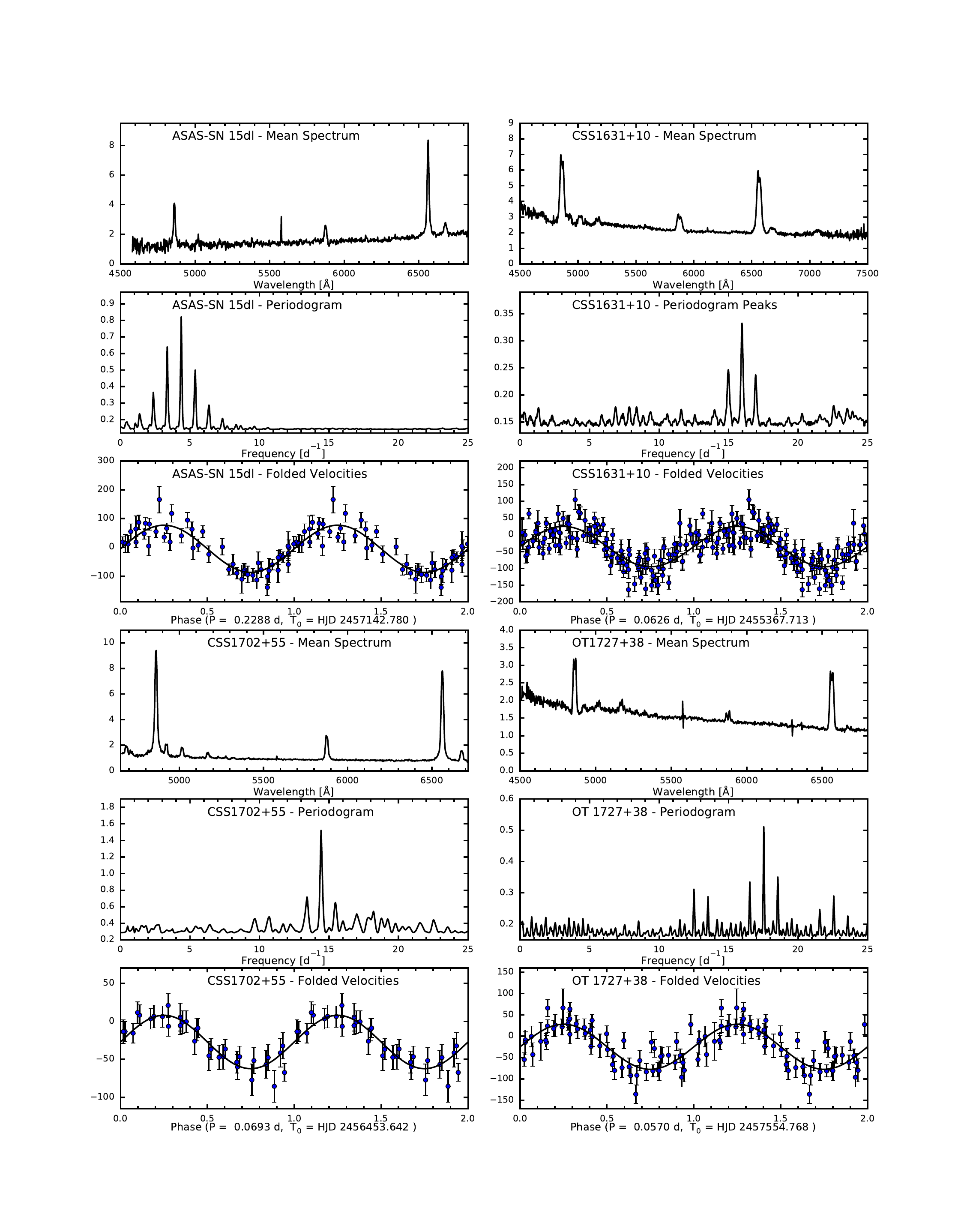}
\caption{Similar to Fig.~\ref{fig:cvplot1}, but for 
ASAS-SN 15dl, CSS1631+10, CSS1702+55, and OT1727+38. 
}
\label{fig:cvplot6}
\end{figure}

\begin{figure}
\includegraphics[height=21.5 cm,trim = 2.2cm 2.0cm 2cm 2.8cm,clip=true]{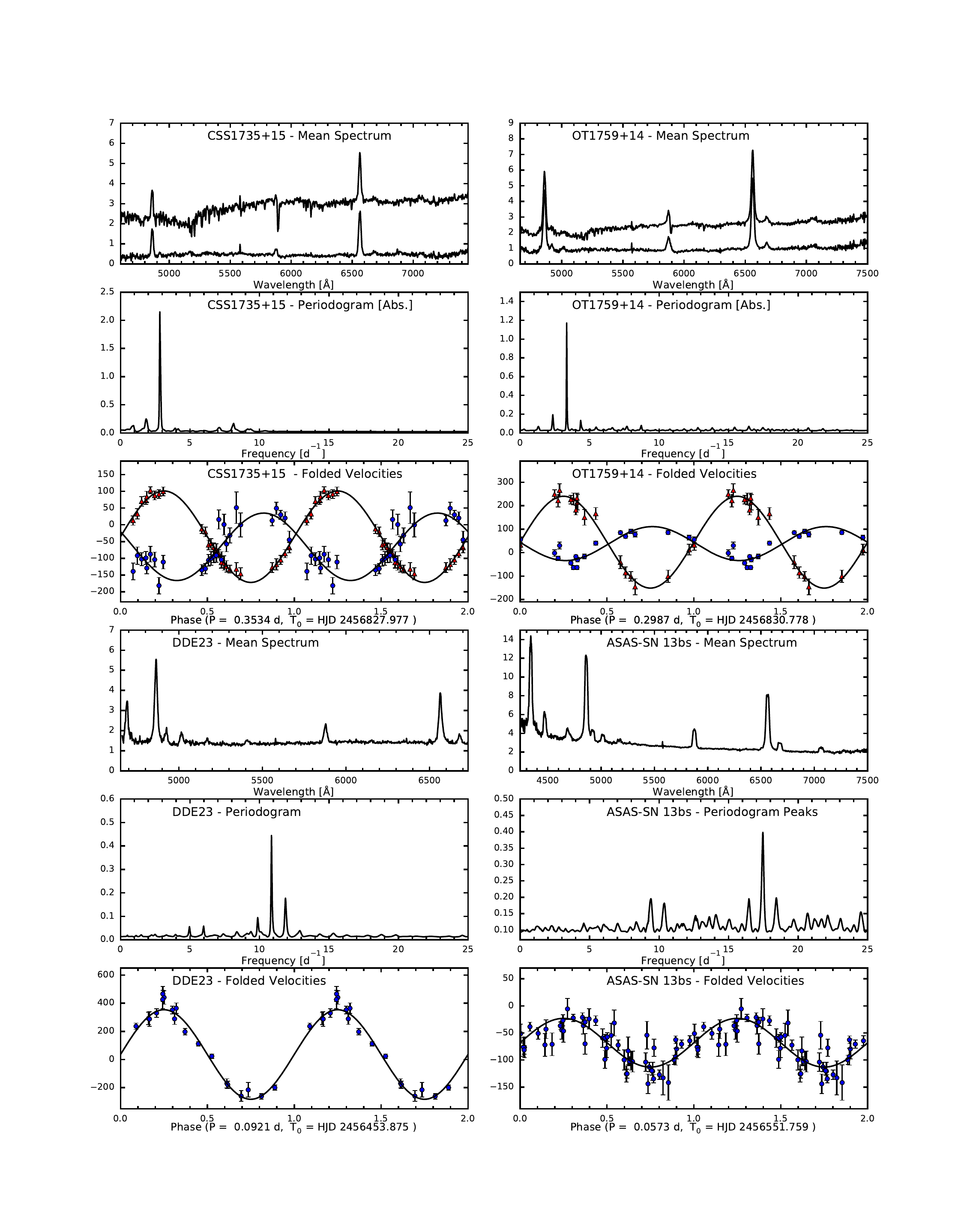}
\caption{Similar to Fig.~\ref{fig:cvplot1}, but for 
CSS1735+15, OT1759+14, DDE23, and ASAS-SN 13bs.  
}
\label{fig:cvplot7}
\end{figure}

\begin{figure}
\includegraphics[height=21.5 cm,trim = 2.2cm 2.0cm 2cm 2.8cm,clip=true]{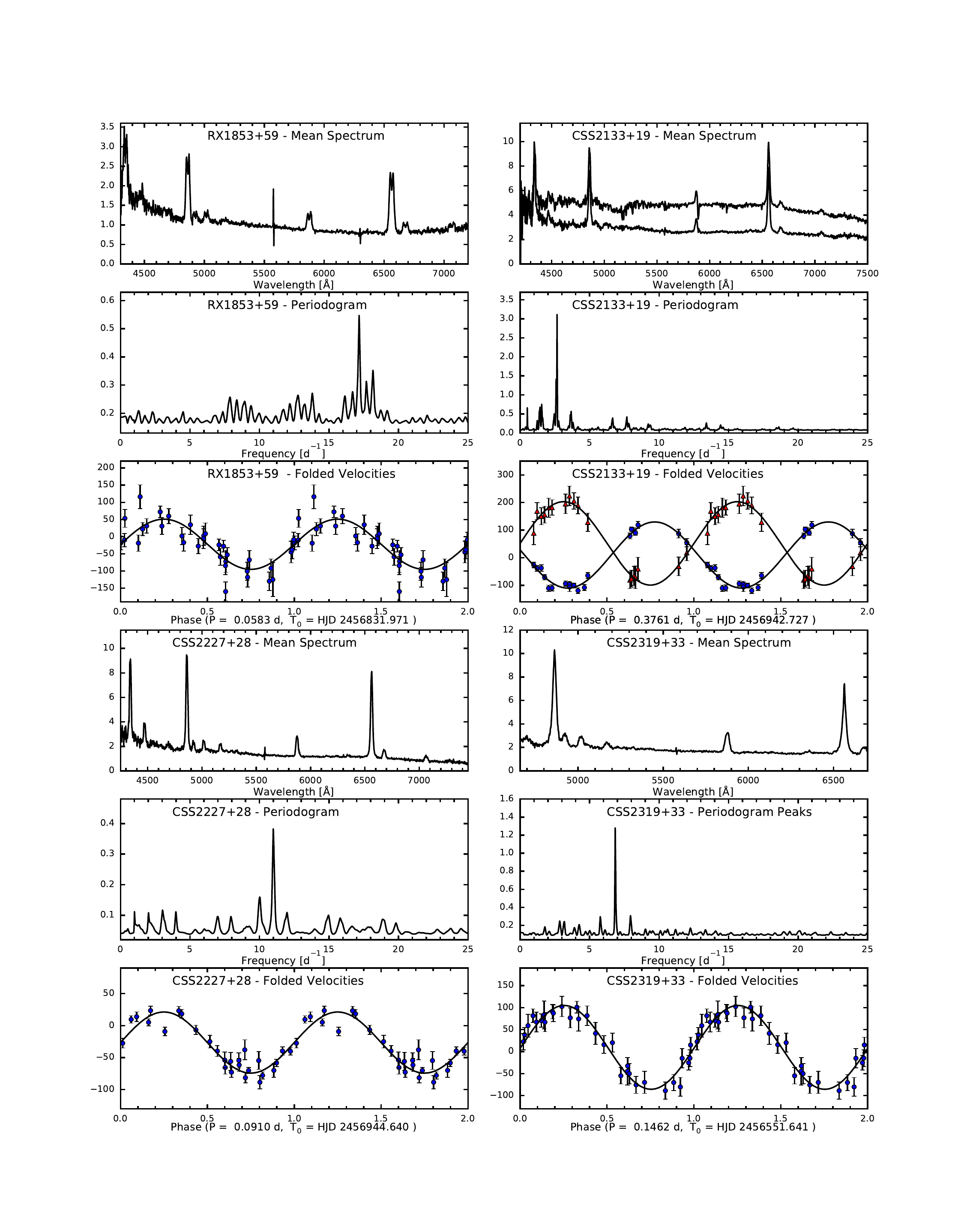}
\caption{Similar to Fig.~\ref{fig:cvplot1}, but for 
RX1853+59,  CSS2133+19, CSS2227+28, and CSS2319+33. 
}
\label{fig:cvplot8}
\end{figure}
 
\begin{figure}
\includegraphics[height=20 cm,trim = 2.2cm 1.0cm 2cm 2.8cm,clip=true]{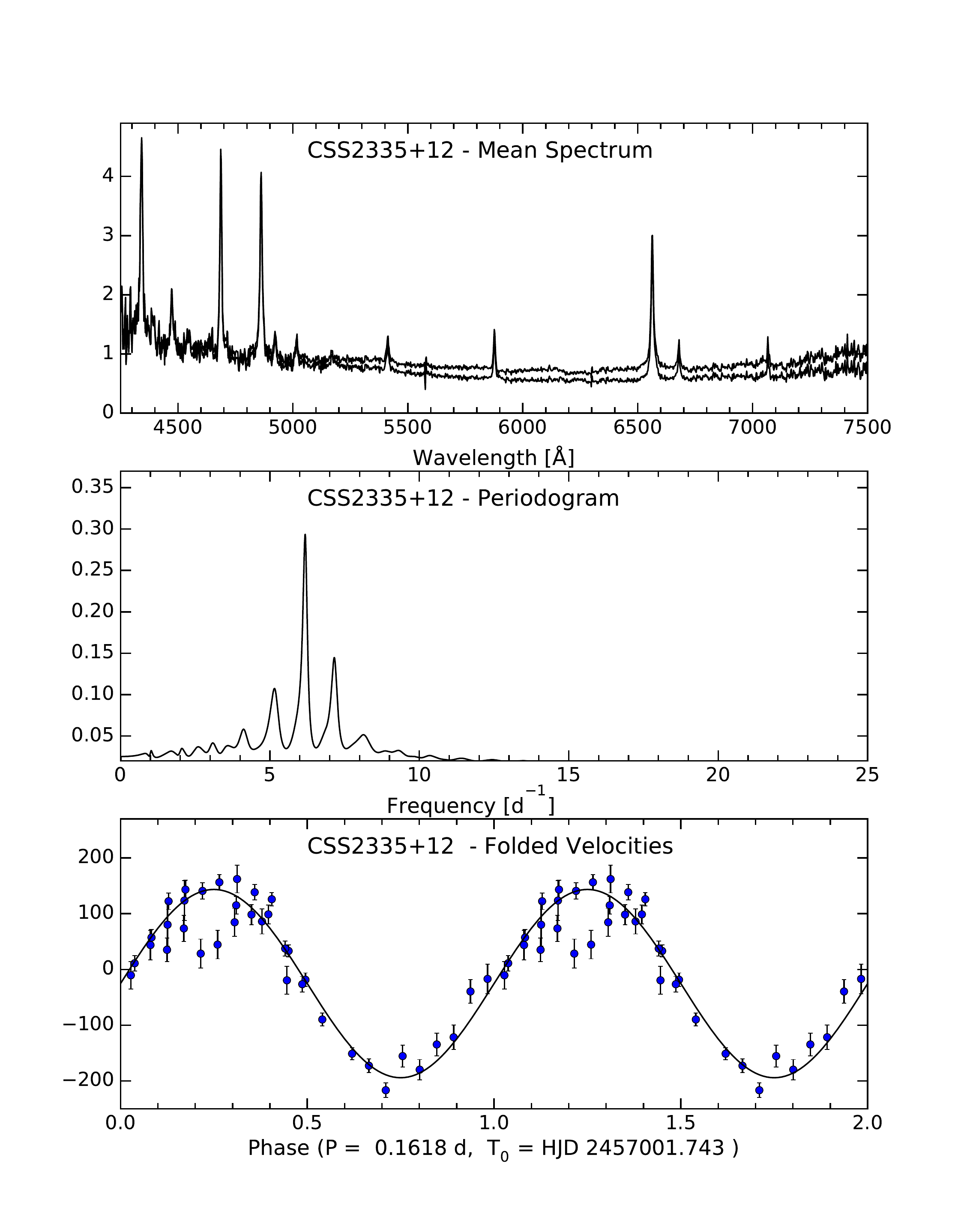}
\caption{Similar to Fig.~\ref{fig:cvplot1}, but for 
CSS2335+12.  
}
\label{fig:cvplot9}
\end{figure}
  
\begin{figure}
\includegraphics[height=13 cm,trim=2cm 0cm 0cm 1.5cm]{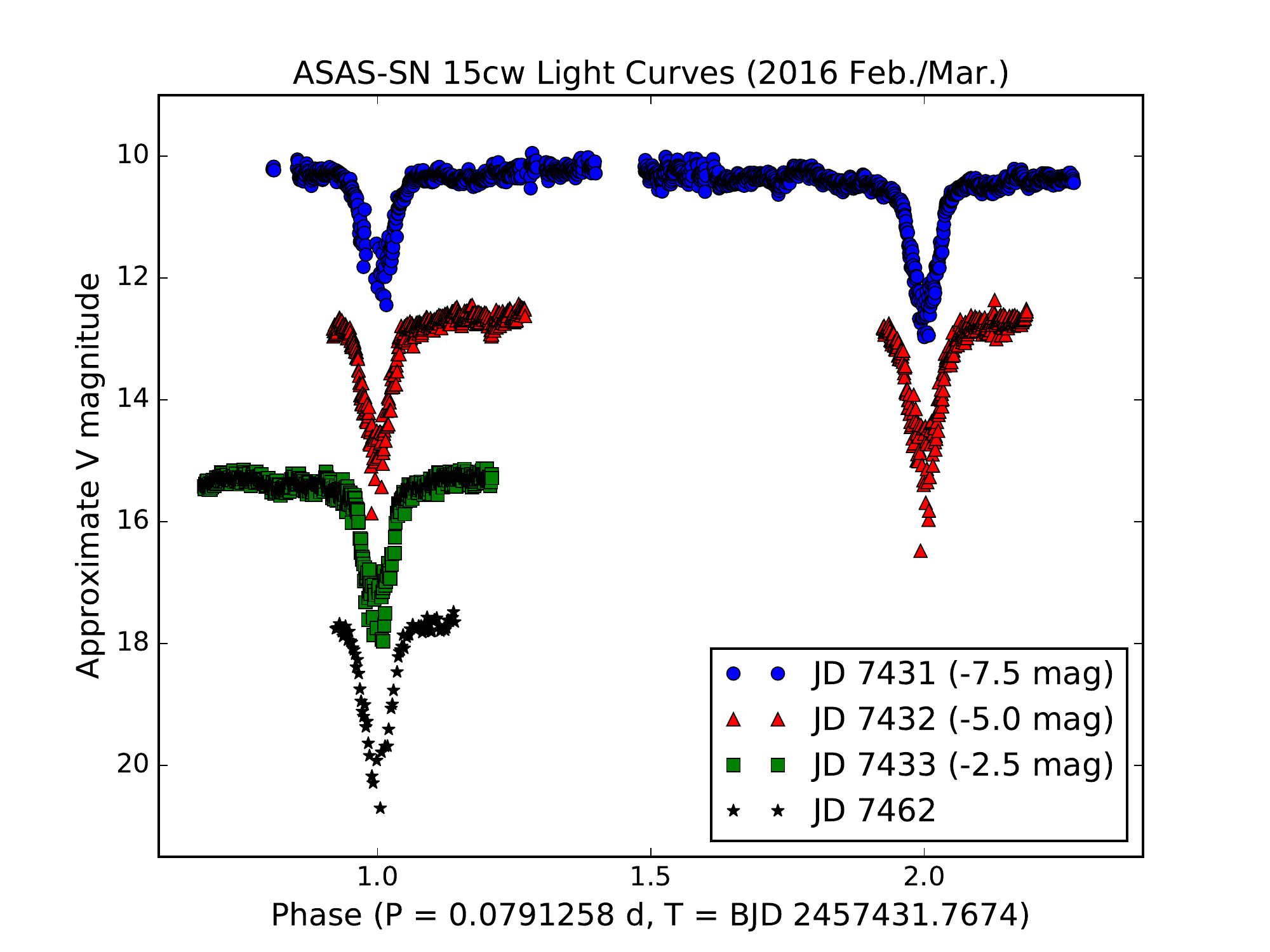}
\caption{Light curves of ASAS-SN 15cw taken in 2016 February and March.
The horizontal axis shows the phase on the indicated ephemeris, and
each night is offset by 2.5 mag The vertical axis shows 
the differential magnitude in white light, adjusted to rough $V$ magnitude 
by adding the magnitude of the comparison star at
$\alpha = $08:08:21.37, $\delta = $00:59:15.7, 
$39''$ from the program object in position angle 67 degrees, 
for which APASS 
\citep{henden15} gives $V = 15.67$.
}
\label{fig:15cweclipse}
\end{figure}

\begin{figure}
\includegraphics[height=18cm,trim=2cm 1cm 0cm 1cm]{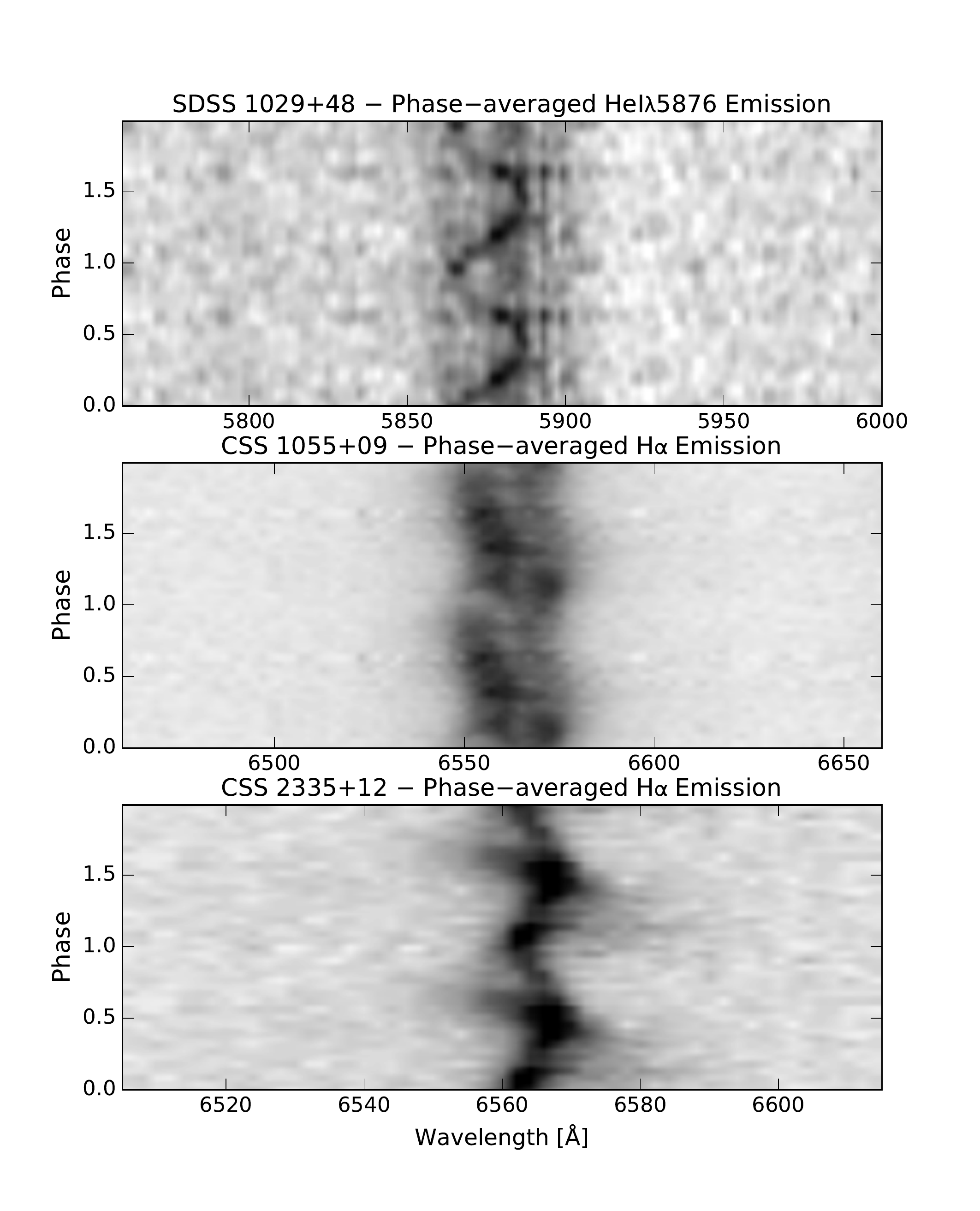}
\caption{(Top panel).  Phase averaged greyscale 
representation of the HeI $\lambda$5876 emission line in 
CSS 1029+48.  The color map
is inverted (emission appears dark).  Each line of the 
image is a weighted average of spectra taken near the
appropriate phase.  Before averaging, the individual 
spectra were rectified (i.e., divided by a fitted 
continuum), and bad pixels were edited out.
The data are repeated for a second cycle.
A clear $S$-wave is seen, and NaD lines at 
constant velocity are just discernible.
(Center panel.) Similar, for H$\alpha$ emission
in CSS1055+09.
(Bottom panel.) Similar, for 
H$\alpha$ emission in CSS 2335+19.}
\label{fig:trails}
\end{figure}

\begin{figure}
\includegraphics[height=18 cm,trim=2cm 0cm 0cm 1.5cm]{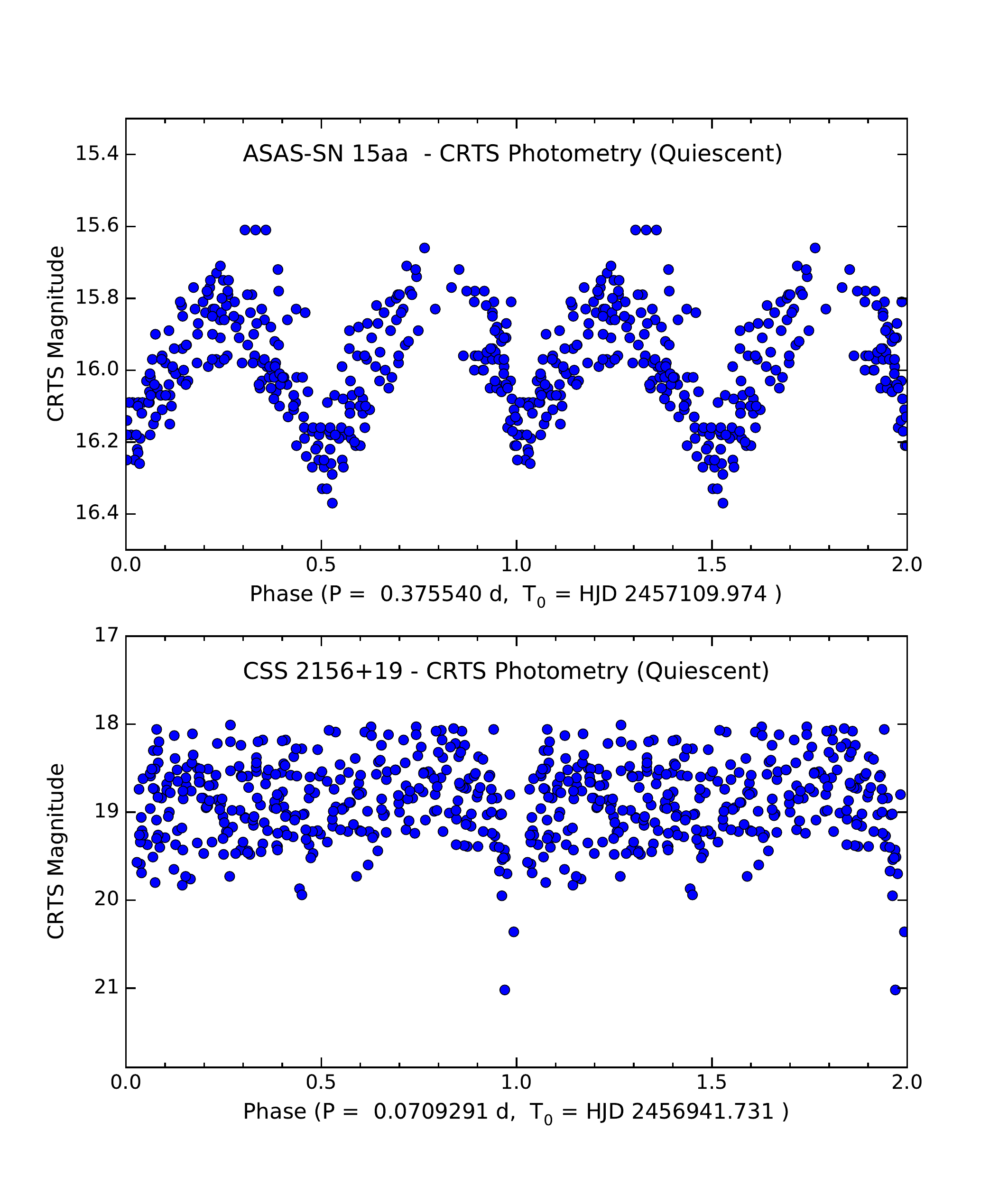}
\caption{CRTS photometry of ASAS-SN 15aa (upper panel) and 
CSS2156+19 (lower panel) folded on candidate long-term periods.
Points taken during outbursts are omitted.
}
\label{fig:crtsphot}
\end{figure}

\clearpage

\begin{figure}
\includegraphics[height=13 cm,trim=2cm 0cm 0cm 1.5cm]{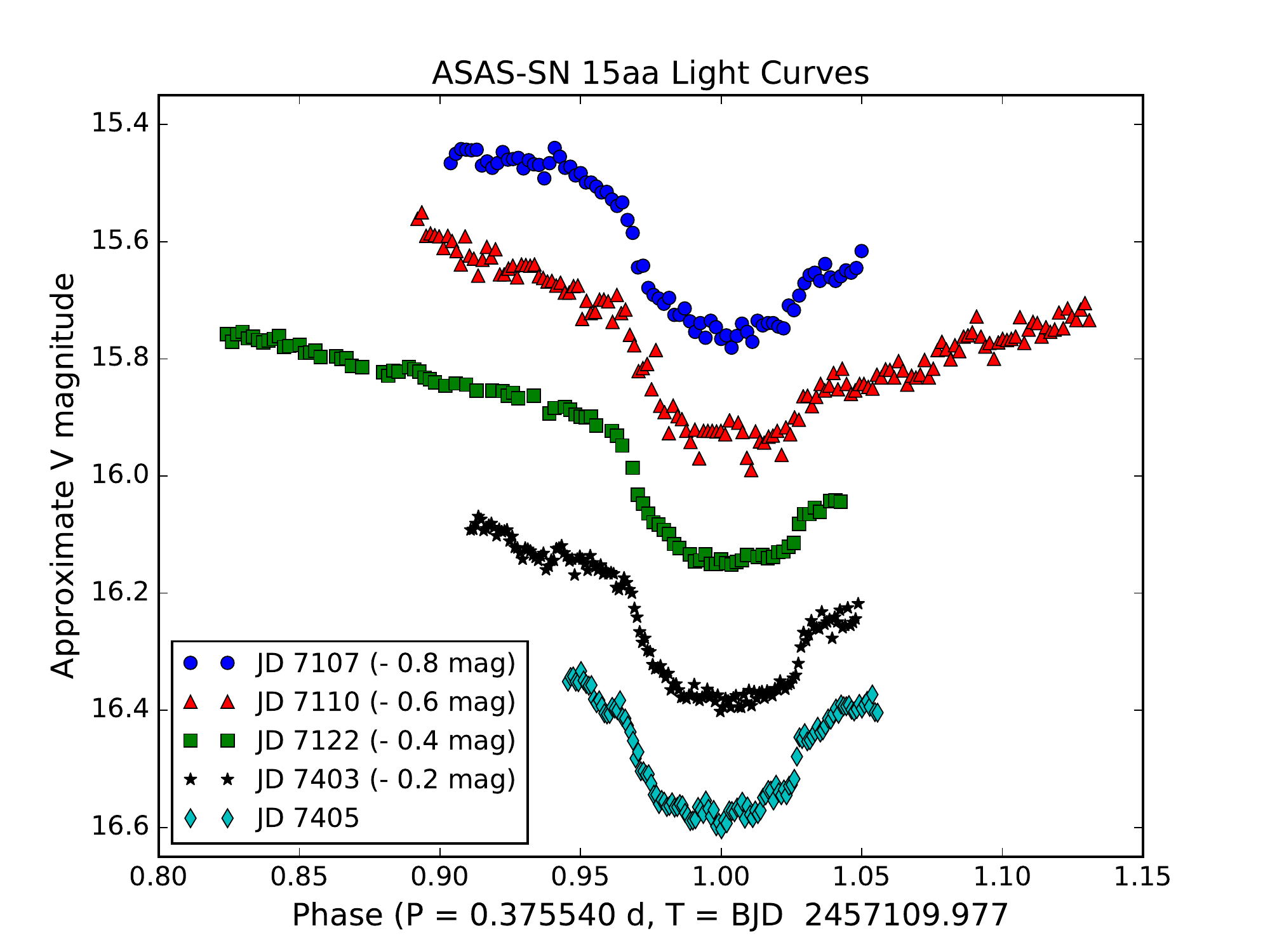}
\caption{Light curves of ASAS-SN 15aa taken in 2015 March and April,
and 2016 January.
The horizontal axis shows the phase on the indicated ephemeris, and
each night is offset by 0.2 magnitude.  The vertical axis shows 
unfiltered differential magnitudes, scaled to rough $V$ magnitude 
by adding the magnitude of the comparison star $88''$ NW of
the program object, at
$\alpha = $10:49:21.02, $\delta = -$21:46:39.6, for which APASS 
\citep{henden15} gives $V = 16.15$.  The gradual fading
preceding the eclipse and the gradual brightening following the
eclipse are probably due to ellipsoidal variation.
}
\label{fig:15aaeclipse}
\end{figure}

\begin{figure}
\includegraphics[height=18cm,trim=2cm 1cm 0cm 1cm]{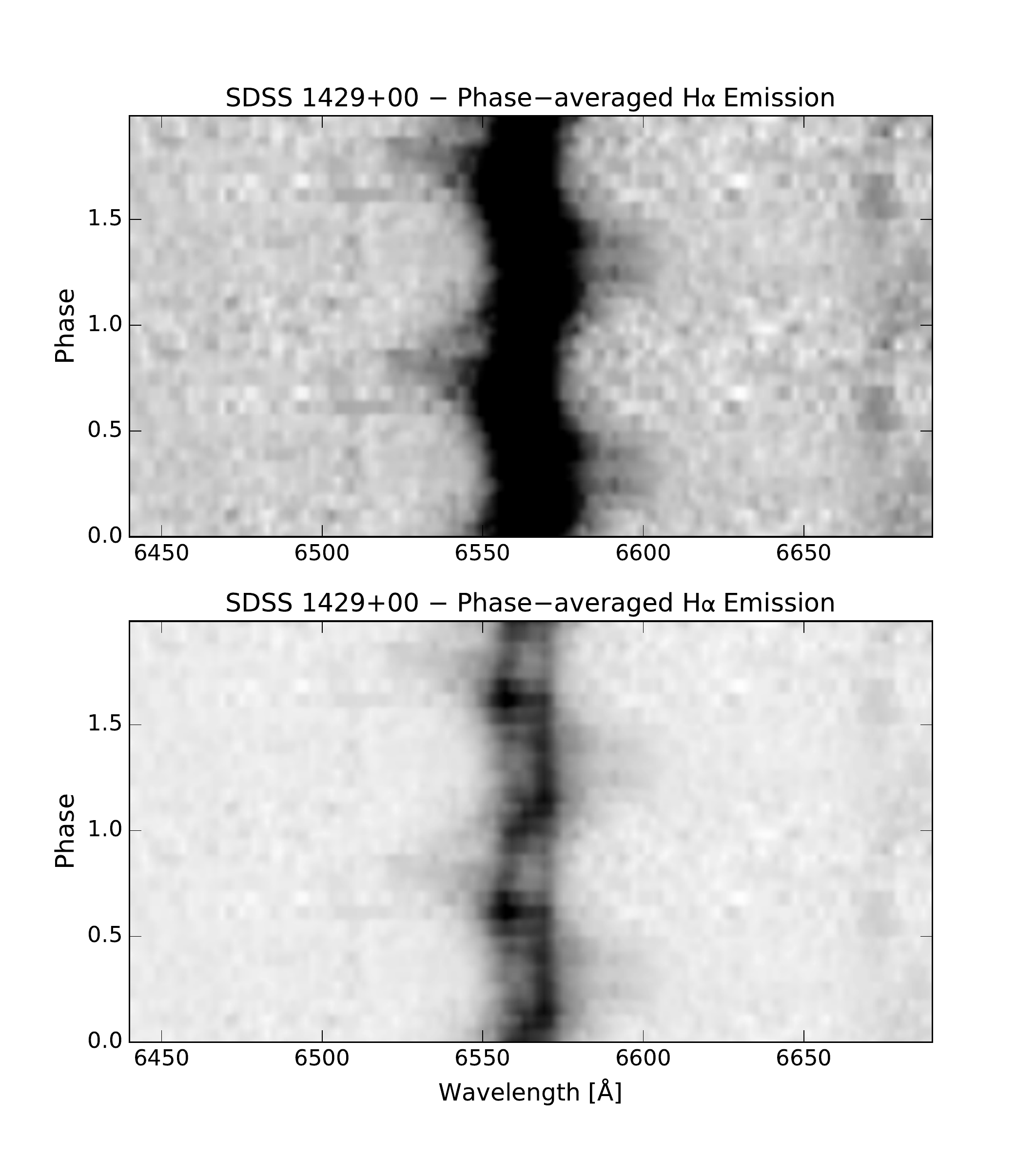}
\caption{(Upper panel).  Phase averaged greyscale 
representation of H$\alpha$ in SDSS 1429+00, prepared
similarly to Fig.~\ref{fig:trails}.  The stretch is 
chosen to emhpasize the high-amplitude $S$-wave.
(Lower panel).  The same data, stretched to show the
line core.
}
\label{fig:sdss1429trail}
\end{figure}

\clearpage

\begin{figure}
\includegraphics[height=18cm,trim=2cm 1cm 0cm 1cm]{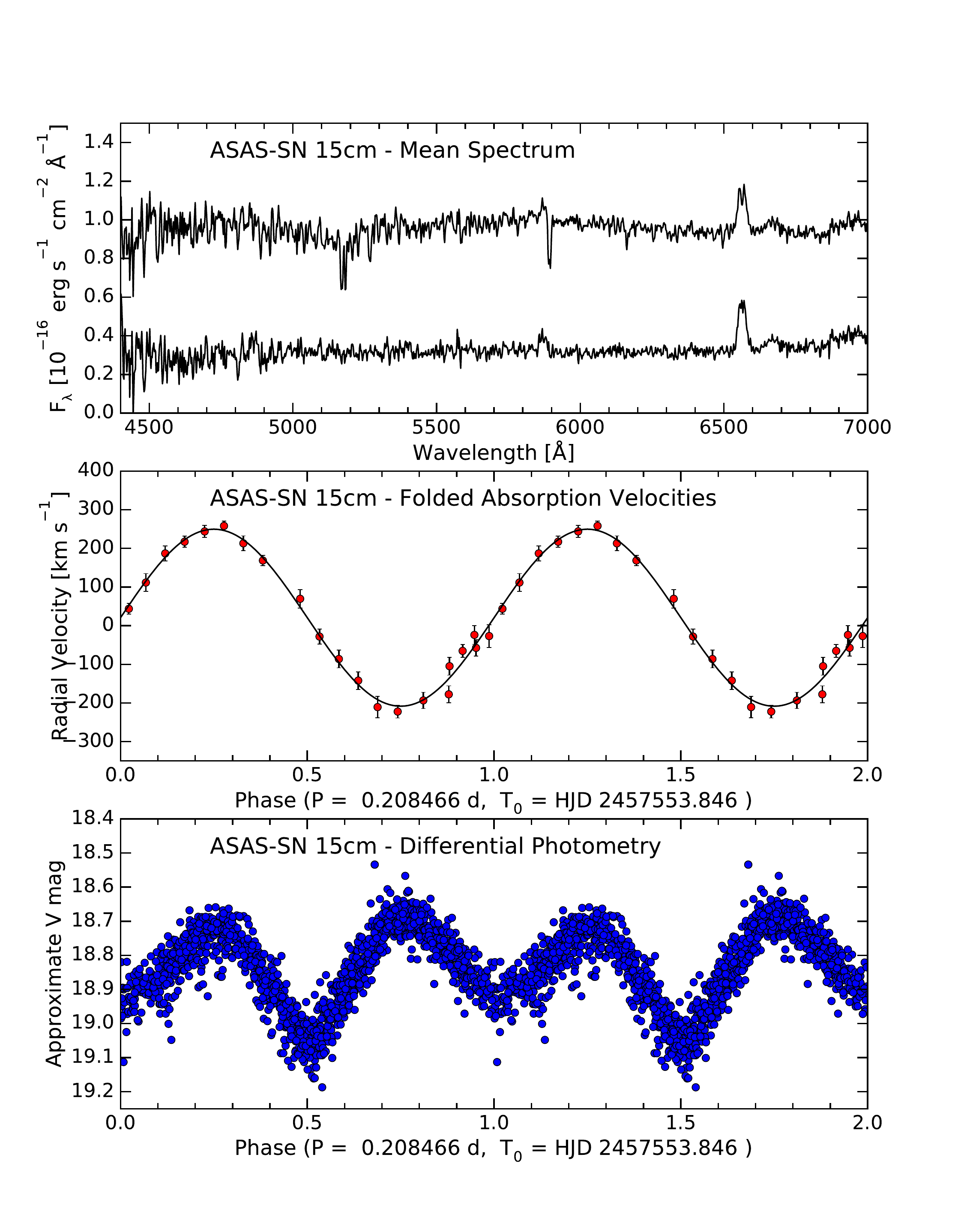}
\caption{(Upper panel). Mean spectrum of ASAS-SN 15cm, formed
by shifting all spectra to the velocity of the secondary star
before averaging.  The lower trace shows the spectrum 
after subtracting a spectrum of the K2V star HD101011, scaled
to a synthetic $V = 19.5$. 
(Center panel.) Absorption-line radial
velocities folded on the orbital period, with the best-fitting
sinusoid (Table~\ref{tab:parameters}) superposed.  (Lower panel.)
White-light differential photometry, folded on the orbital 
period.  The main comparison star lies 23 arcsec from 
the target in PA 155 degrees; its SDSS magnitudes imply
$V \sim 16.1$.  The vertical scale has been offset by this
amount to give approximate $V$ magnitudes.
}
\label{fig:asn15cm}
\end{figure}

\clearpage

\begin{figure}
\includegraphics[height=13 cm,trim=2cm 0cm 0cm 1.5cm]{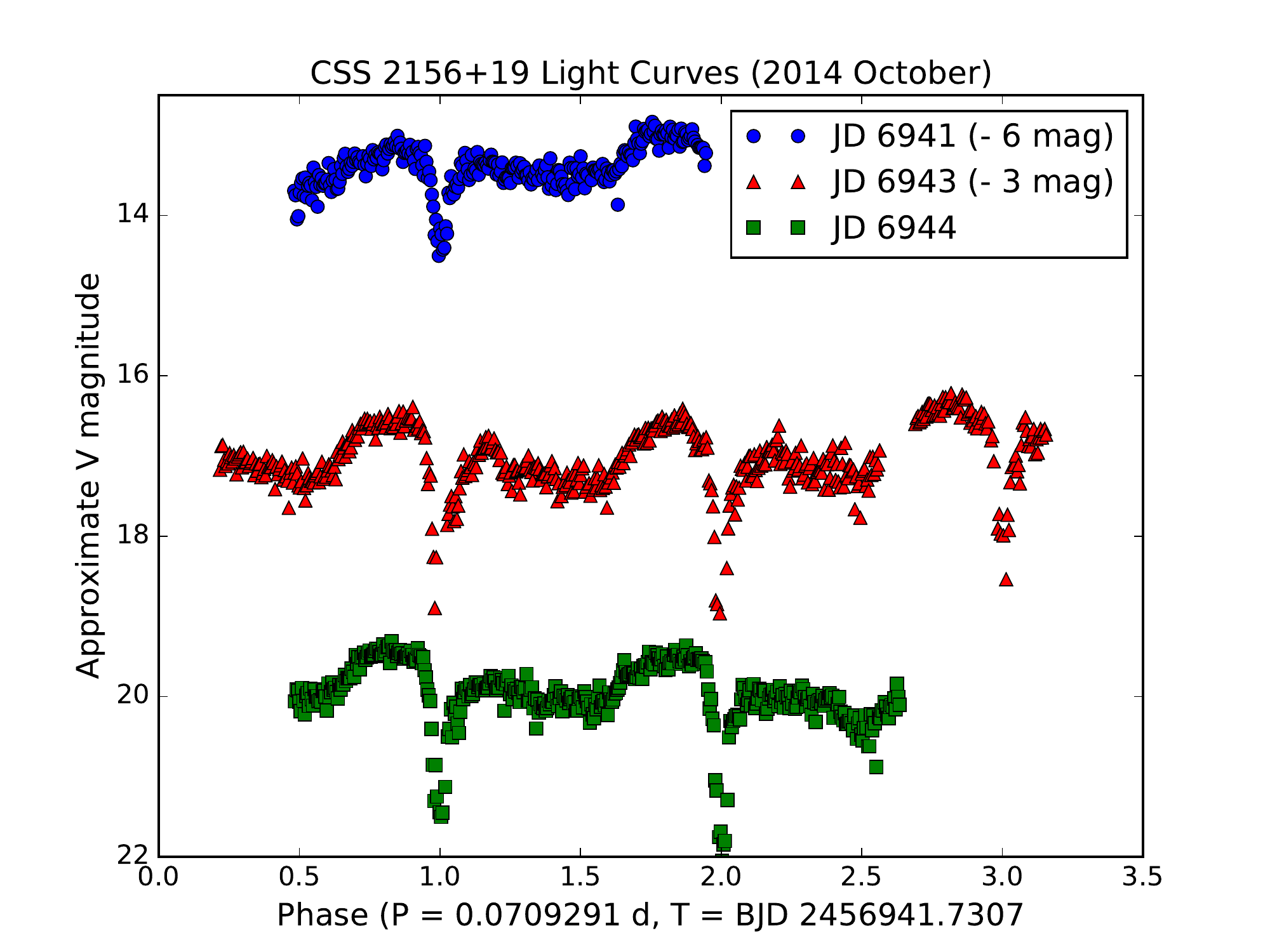}
\caption{Light curves of CSS2156+19 taken on three nights in 
2014 October.  The horizontal axis shows phase on the indicated
ephemeris, with phase increasing monotonically through each night,
and nights offset by 3 magnitudes as indicated.  The $V$ magnitude
used for the vertical scale is approximate, since the data 
were taken in white light 
(GG420 long-pass filter).  The 
magnitudes are differential with respect to a star at 
$\alpha = $21:56:38.472, $\delta = +$19:33:17.96, for 
which we assumed $V = 16.18$ based on SDSS DR12.
}
\label{fig:css2156lightcurves}
\end{figure}

\clearpage

\begin{figure}
\includegraphics[height=12 cm,trim=2cm 0cm 0cm 1.5cm]{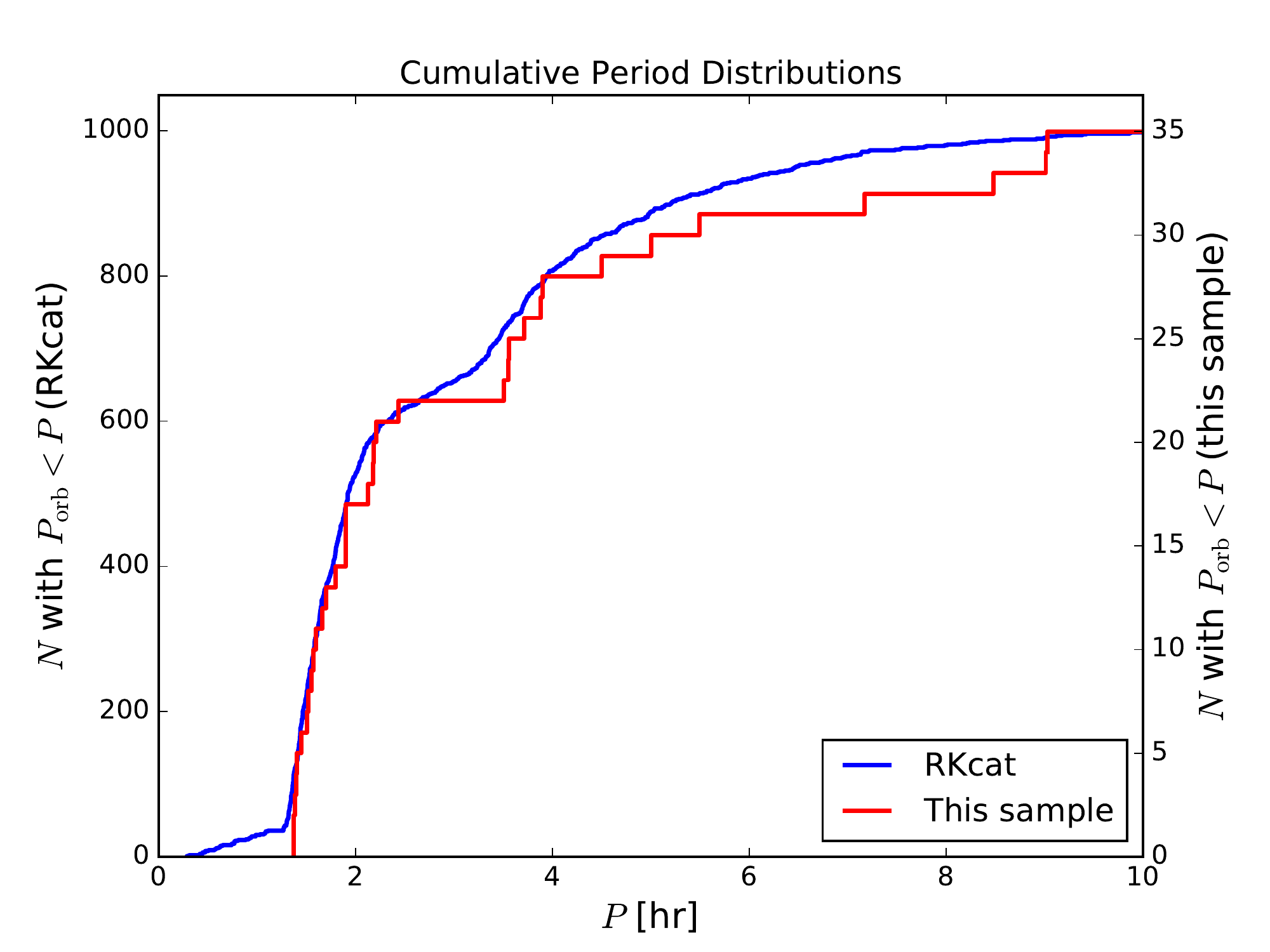}
\caption{Cumulative distribution function for the present 
sample of 35 CVs, compared with the sample of CVs from
the Ritter-Kolb catalog (restricted to $P_{\rm orb} < 10$ h).
} 
\label{fig:cdf}
\end{figure}


\clearpage

\begin{thebibliography}

\bibitem[Baraffe \& Kolb(2000)]{bk00} Baraffe, I., Kolb, U. 2000,
\mnras, 318, 354

\bibitem[Breedt et al.(2014)]{breedt14} Breedt, E.,
G{\"a}nsicke, B.~T., Drake, A.~J., et al.\ 2014, \mnras, 443, 3174

\bibitem[Carter et al.(2013)]{carter13} Carter, P.~J., 
Marsh, T.~R., Steeghs, D., et al.\ 2013, \mnras, 429, 2143 

\bibitem[Chanan et al.(1976)]{chanan76} Chanan, G.~A., 
Middleditch, J., \& Nelson, J.~E.\ 1976, \apj, 208, 512 

\bibitem[Denisenko \& Sokolovsky(2011)]{denisenko11} Denisenko, D.~V., 
\& Sokolovsky, K.~V.\ 2011, Astronomy Letters, 37, 91 

\bibitem[Denisenko(2012)]{denisenko23} Denisenko, D.~V.\ 2012, 
Astronomy Letters, 38, 249 


\bibitem[Denisenko et al.(2014)]{denisenko14} Denisenko, D., Lipunov, V., Gorbovskoy, E., et al.\ 2014, The Astronomer's Telegram, 5724  

\bibitem[Drake et al.(2009)]{drakecrtts} Drake, A.~J., Djorgovski, 
S.~G., Mahabal, A., et al.\ 2009, \apj, 696, 870 

\bibitem[Drake et al.(2014a)]{drakevars14} Drake, A.~J., Graham, 
M.~J., Djorgovski, S.~G., et al.\ 2014, \apjs, 213, 9 

\bibitem[Drake et al.(2014b)]{drake14} Drake, A.~J., 
G{\"a}nsicke, B.~T., Djorgovski, S.~G., et al.\ 2014, \mnras, 441, 1186 

\bibitem[G{\"a}nsicke et al.(2009)]{unveils} G{\"a}nsicke, 
B.~T., Dillon, M., Southworth, J., et al.\ 2009, \mnras, 397, 2170 

\bibitem[Filippenko(1982)]{filippenko82} Filippenko, A.~V.\ 1982, 
\pasp, 94, 715 

\bibitem[Goliasch \& Nelson(2015)]{goliaschnelson15} Goliasch, J., 
\& Nelson, L.\ 2015, \apj, 809, 80 

\bibitem[Green et al.(1982)]{green82} Green, R.~F., Ferguson, 
D.~H., Liebert, J., \& Schmidt, M.\ 1982, \pasp, 94, 560 

\bibitem[Haakonsen \& Rutledge(2009)]{haakonsen09} Haakonsen, C.~B., \& Rutledge, 
R.~E.\ 2009, \apjs, 184, 138 

\bibitem[Halpern \& Thorstensen(2015)]{halpernthor15} Halpern, J.~P., 
\& Thorstensen, J.~R.\ 2015, \aj, 150, 170 

\bibitem[Haswell et al.(1994)]{haswell94} Haswell, C.~A., Horne, K., 
Thomas, G., Patterson, J., \& Thorstensen, J.~R.\ 1994, Interacting Binary Stars, 56, 268 

\bibitem[Henden et al.(2015)]{henden15} Henden, A.~A., Levine, 
S., Terrell, D., \& Welch, D.~L.\ 2015, American Astronomical Society 
Meeting Abstracts, 225, \#336.16 

\bibitem[Kato et al.(2010)]{kato_ii} Kato, T., Maehara, H., 
Uemura, M., et al.\ 2010, \pasj, 62, 1525 

\bibitem[Kato et al.(2013)]{kato_iv} Kato, T., Hambsch, F.-J.,
Maehara, H., et al.\ 2013, \pasj, 65, 23

\bibitem[Kato et al.(2014)]{kato14} Kato, T., Hambsch, F.-J., 
Maehara, H., et al.\ 2014, \pasj, 66, 30 

\bibitem[Kato et al.(2015)]{katovii} Kato, T., Hambsch, F.-J., 
Dubovsky, P.~A., et al.\ 2015, \pasj, 67, 105 

\bibitem[Kaur et al.(2014)]{kaur14} Kaur, A., Porter, A., 
Wilber, A., et al.\ 2014, The Astronomer's Telegram, 6624, 1 

\bibitem[Kepler et al.(2016)]{kepler16} Kepler, S.~O., Pelisoli, I., 
Koester, D., et al.\ 2016, \mnras, 455, 3413 

\bibitem[Knigge(2006)]{kniggedonors} Knigge, C.\ 2006, \mnras, 373, 
484 

\bibitem[Kurtz \& Mink(1998)]{kurtzmink} Kurtz, M.~J. \& Mink, 
D.~J.\ 1998, \pasp, 110, 934

\bibitem[Lazavera et al.(2013)]{lazavera13} Lazavera, A., Voroshilov, N., 
Denisenko, D., Kuznetsov, A., Gorbovskoy, E., \& Lipunov, V. 2013, 
arXiv:1307:6855    

\bibitem[Lipunov et al.(2010)]{lipunov10} Lipunov, V., Kornilov, 
V., Gorbovskoy, E., et al.\ 2010, Advances in Astronomy, 2010, 349171 

\bibitem[McLaughlin(1924)]{mclaughlin24} McLaughlin, D.~B.\ 1924, \apj, 60, 22 

\bibitem[Monet et al.(2003)]{usnob} Monet, D.~G., Levine, 
S.~E., Canzian, B., et al.\ 2003, \aj, 125, 984 

\bibitem[Patterson(1998)]{pattlate} Patterson, J.\ 1998, \pasp, 110, 1132

\bibitem[Rebassa-Mansergas et al.(2014)]{rebassa14} 
Rebassa-Mansergas, A., Parsons, S.~G., Copperwheat, 
C.~M., et al.\ 2014, \apj, 790, 28 

\bibitem[Ringwald(1993)]{ringwald93} Ringwald, F.~A.\ 1993, \pasp, 
105, 805 

\bibitem[Ritter \& Kolb(2003)]{rkcat} Ritter, H., \& Kolb, U.\ 2003, \aap, 
404, 301 

\bibitem[Roeser et al.(2010)]{ppmxl} Roeser, S., Demleitner, 
M., \& Schilbach, E.\ 2010, \aj, 139, 2440 

\bibitem[Rossiter(1924)]{rossiter24} Rossiter, R.~A.\ 1924, \apj, 60,  

\bibitem[Schlegel, Finkbeiner, \& Davis(1998)]{schlegel98}
Schlegel, D. J., Finkbeiner, D. P., \& Davis, M. 1998, \apj, 500, 525
2440 

\bibitem[Schneider \& Young(1980)]{sy80} Schneider, D. and Young, P. 1980,
\apj, 238, 946

\bibitem[Schwope \& Thinius(2012)]{schwope12} Schwope, A.~D., 
\& Thinius, B.\ 2012, Astronomische Nachrichten, 333, 717 

\bibitem[Shappee et al.(2013a)]{shappee13} Shappee, B., Kochanek, 
C.~S., Stanek, K.~Z., et al.\ 2013a, The Astronomer's Telegram, 4987, 1 

\bibitem[Shappee et al.(2014b)]{shappeecurtain} Shappee, B.~J., Prieto, 
J.~L., Grupe, D., et al.\ 2014b, \apj, 788, 48 

\bibitem[Szkody et al.(2011)]{szkodyviii} Szkody, P., Anderson,
S.~F., Brooks, K., et al.\ 2011, \aj, 142, 181

\bibitem[Szkody et al.(2014)]{szkody14} Szkody, P., Everett, 
M.~E., Howell, S.~B., et al.\ 2014, \aj, 148, 63  

\bibitem[Thorstensen(2003)]{thorparallax03} Thorstensen, J.~R.\ 2003, 
\aj, 126, 3017 

\bibitem[Thorstensen \& Freed(1985)]{tf85} Thorstensen, J.~R., 
\& Freed, I.~W.\ 1985, \aj, 90, 2082 

\bibitem[Thorstensen et al.(1991)]{thorstensen91} Thorstensen, J.~R., 
Ringwald, F.~A., Wade, R.~A., Schmidt, G.~D., 
\& Norsworthy, J.~E.\ 1991, \aj, 102, 272 

\bibitem[Thorstensen(2015)]{asn13cl} Thorstensen, J.~R.\ 2015, 
\pasp, 127, 351 

\bibitem[Thorstensen \& Armstrong(2005)]{TA05} Thorstensen, J.~R., 
\& Armstrong, E.\ 2005, \aj, 130, 759 

\bibitem[Thorstensen \& Skinner(2012)]{thorskinner} Thorstensen, J.~R., 
\& Skinner, J.~N.\ 2012, \aj, 144, 81  

\bibitem[Thorstensen \& Halpern(2013)]{thorhalpern13} Thorstensen, 
J.~R., \& Halpern, J.\ 2013, \aj, 146, 107 

\bibitem[Thorstensen(2013)]{thorcss1340} Thorstensen, J.~R.\ 2013, \pasp, 125, 506 

\bibitem[Thorstensen(2015)]{thorasn13cl} Thorstensen, J.~R.\ 2015, \pasp, 127, 351 

\bibitem[Tiurina et al.(2013)]{tiurina13} Tiurina, N., Frolova, 
N., Pruzghinskaya, M., et al.\ 2013, The Astronomer's Telegram, 4871, 1 

\bibitem[Tonry \& Davis(1979)]{tonrydavis} Tonry, J., \& Davis, M.\ 1979, \aj, 84, 1511 

\bibitem[Warner(1987)]{warn87} Warner, B.\ 1987, \mnras, 227, 23

\bibitem[Warner(1995)]{warner95} Warner, B., in {\it Cataclysmic Variable
Stars}, 1995, Cambridge University Press, New York

\bibitem[West et al.(2008)]{west08} West, A.~A., Hawley, S.~L., Bochanski, J.~J., et al.\ 2008, \aj, 135, 785 

\bibitem[Zacharias et al.(2015)]{urat} Zacharias, N., Finch, 
C., Subasavage, J., et al.\ 2015, VizieR Online Data Catalog, 1329, 0

\bibitem[Zharikov et al.(2013)]{zharikov13} Zharikov, S., Tovmassian, G., Aviles, A., et al.\ 2013, \aap, 549, A77 




\end{thebibliography}
\end{document}